\documentclass[a4paper,fleqn,usenatbib]{mnras}
\usepackage{graphicx}	
\usepackage{amsmath}	
\usepackage{amssymb}	
\usepackage{comment}
\usepackage{soul}

\def\msun{{\rm M_{\odot}}}

\def\be{\begin{equation}}
\def\ee{\end{equation}}

\def\del#1{{}}
\def\SNc#1{\textcolor{red}{#1}}
\newcommand{\bref}[1]{{#1}}
\def\YWc#1{\textcolor{green}{#1}}
\newcommand\mearth{{\,{\rm M}_{\oplus}}}
\newcommand\mj{{\,{\rm M}_{\rm J}}}
\newcommand\St{\,{\rm St}}

\title[]{Gap opening by planets in discs with magnetised winds}

\author[Elbakyan et al.]{Vardan Elbakyan$^1$\thanks{ve23@le.ac.uk}, Yinhao Wu$^{1}$,  Sergei Nayakshin$^{1}$, Giovanni Rosotti$^{1}$\\
$^{1}$School of Physics and Astronomy, University of
  Leicester, Leicester LE1 7RH, UK. 
}

\date{Accepted XXX. Received YYY; in original form ZZZ}

\pubyear{2022}

\begin{document}
\label{firstpage}
\pagerange{\pageref{firstpage}--\pageref{lastpage}}
\maketitle

\begin{abstract}
Planets open deep gaps in protoplanetary discs when their mass exceeds a gap opening mass, $M_{\rm gap}$. We use one- and two-dimensional simulations to study planet gap opening in discs with angular momentum transport powered by  MHD disc winds. We parameterise the efficiency of the MHD disc wind angular momentum transport through a dimensionless parameter $\alpha_{\rm dw}$, which is an analogue to the turbulent viscosity $\alpha_{\rm v}$. We find that magnetised winds are much less efficient in counteracting planet tidal torques than turbulence is. \bref{For discs with astrophysically realistic values of $\alpha_{\rm dw}$, $M_{\rm gap}$ is always determined by the residual disc turbulence, and is a factor of a few to ten smaller than usually obtained for viscous discs.} We introduce a gap opening criterion applicable for any values of $\alpha_{\rm v}$ and $\alpha_{\rm dw}$ \bref{that may be useful for planet formation population synthesis. We show that in discs powered by magnetised winds growing planets detach from the disc at planet masses below $\sim 0.1\mj$ inside 10 AU. This promotes formation of super-Earth planets rather than gas giants in this region, in particular precluding formation of hot jupiters in situ. On larger scales, ALMA gap opening planet candidates may be less massive than currently believed.} Future high-resolution observations with instruments such as the extended ALMA, ngVLA, and SKA are likely to show abundant narrow annular features at $R < 10$~AU  due to ubiquitous super-Earth planets.



\end{abstract}


\section{Introduction}

A gas giant planet embedded in a protoplanetary disc produces tidal torques that tend to push gas away from its vicinity, carving a gap in the disc at sufficiently high planet masses \citep[e.g.,][]{GoldreichTremaine80,LinPap86,BaruteauEtal14a}. Planet mediated gap opening in the dust component of protoplanetary discs is even more efficient \citep[][]{DipierroEtal15,RosottiEtal16,DipierroLaibe17}, requiring lower planet masses. These theoretical predictions have helped to infer a widespread presence of planets in a couple dozen bright discs resolved with ALMA \citep[e.g.,][]{BroganEtal15,HuangEtal16,2018HuangAndrews,2018LongPinilla}. 

Gap opening by planets embedded in protoplanetary discs has a significant impact on a wide range of phenomena in star and planet formation. The maximum mass to which a gas giant planet will grow via gas accretion runaway is established when the planet isolates itself from the gas  supply \citep[][]{Bate03,IdaLin04a,MordasiniEtal09b,AyliffeBate09}. The rate and direction with which the planet migrates in the disk depends on whether it does so in the type I \citep[no deep gap opened;][]{Tanaka02,Paardekooper11-typeI} or in the type II regime \citep[e.g.,][]{2012KleyNelson,BaruteauEtal14a}. That in turn determines whether the planet migrates to the inner edge of the disc or even into the star, or stops at a much larger radius \citep[][]{AA09,ColemanNelson14}. Planets migrating in the type II regime can filter dust particles on the gap edges \citep[e.g.,][]{RiceEtal03b,ZhuEtal12b} and also interrupt gas supply onto the star, which can be important for FU Ori outbursts of young stars \citep[e.g.,][]{LodatoClarke04,NayakshinLodato12} and wide inner holes in transition discs \citep[e.g.,][]{ZhuEtal11}, such as the system PDS70 where two gas giants planets are unambiguously detected \citep[e.g.,][]{HaffertEtal19,KepplerEtal19,BenistyEtal21}.

The minimum planet mass needed to open a deep gap in the gas disk, $M_{\rm p}$, 
depends primarily on the stellar mass, disc aspect ratio, $H/R$, and also on the poorly known turbulence (aka viscosity) parameter $\alpha_{\rm v}$ introduced by \cite{Shakura73}. \cite{CridaEtal06} worked out a very widely used condition for a gas disc gap opening that makes this apparent. For a gap with a factor of ten depression in the gas surface density profile at the location of the planet,  the parameter $\mathcal{P}$
\begin{equation}
    \mathcal{P} = \frac{3H}{4R_{\rm H}} + \frac{50 \alpha_{\rm v}}{q} \left({\frac{H}{R}}\right)^2 
\label{CridaP0}
\end{equation}
needs to drop below unity. At $\mathcal{P} > 1$ the planet migrates in the type I regime. Here $q = M_{\rm p}/M_*$, $M_*$ is the mass of the star, and $R_{\rm H} = R (M_{\rm p}/3M_*)^{1/3}$ is the Hill radius. At very low $\alpha_{\rm v} \sim 10^{-4}$, the first term on the right in eq.~(\ref{CridaP0}) dominates, and the gap opening planet-to-star mass ratio is dictated by the thermal disc structure, that is, the value of $h = H/R$:
\begin{equation}
    q_{\rm th}  \approx h^3 = 10^{-3}  \left[\frac{ h}{0.1}\right]^3\;.
\end{equation}  
In the opposite limit, the higher $\alpha_{\rm v}$, the higher is $M_{\rm p}$. Dropping the first term on the right hand side in eq.~(\ref{CridaP0}), we have
\begin{equation}
    q_{\rm visc} =50  \alpha_{\rm v} h^2 =  5 \times 10^{-3} 
    \left[\frac{ \alpha_{\rm v}}{10^{-2}}\right]
     \left[\frac{ h}{0.1}\right]^2\;.
     \label{q_visc}
\end{equation}


Although some authors deduced from the observed disc radial sizes and gas accretion rates onto the stars values of $\alpha_{\rm v}$ up to $10^{-1}$ \citep{2017Rafikov,2018AnsdellWilliams}, more careful analysis of dust evolution in these discs typically leads to $\alpha_{\rm v}\lesssim 10^{-3}$ \citep[][]{2017Lodato, 2018ZhangDSHARP,2020Rosotti,2021DoiKataoka,2022Villenave}. Furthermore, ALMA observations of molecular lines and vertical/radial distribution of dust provide additional ways of measuring turbulent velocities in protoplanetary discs, which can be used to constrain $\alpha_{\rm v}$ \citep[see a recent review in][]{2022PintePPVII}. Such analysis indicates $\alpha_{\rm v}\lesssim 10^{-3}$ for most of the observed discs \citep[][]{2017Lodato, DSHARP-6, 2020Rosotti, 2021DoiKataoka}, with a few exceptional cases (e.g., DM Tau) having $\alpha_{\rm v}\approx10^{-2}-10^{-1}$ \citep{2019OhashiKataoka, 2020Flaherty}. Baring these high outlier values for $\alpha_{\rm v}$, the implied $M_{\rm gap}$ masses would be quite low.

The low levels of turbulence observed by ALMA are at odds with the stirred-up vertical distribution of micron sized dust in some discs, and they are also insufficient to account for the significant mass accretion rates onto the stars \citep[cf. the review and references in][]{2022MiotelloPPVII}. Thus, a mechanism that transports mass and angular momentum efficiently and lifts up the micron dust without introducing significant turbulent motions is needed. A possible mechanism responsible for such "stirred, not shaken" discs could be the MHD driven wind \citep[e.g.,][]{2013BaiStone}. A number of authors have recently argued that a weak net flux of vertical magnetic field present in molecular cloud cores may enhance MHD-driven disc winds that remove angular momentum of gas at a faster rate than disc turbulence
\citep[e.g.,][]{Armitage13-MHD-wind, Suzuki16-MHD-winds, Hasegawa17-MHD-wind, Lesur21-MHD-winds, Tabone22-general}. The winds capable of angular momentum removal and lifting dust grains into the disc atmosphere are common in Class 0/I discs, and are also observed in a few Class II sources \citep[see a review by][]{2022PascucciPPVII}.



The goal of our paper is to evaluate the gap opening planet mass in discs where mass and angular momentum transfer are dominated by magnetised winds rather than disc turbulence.
 
The paper is organised as follows. In Sect.~\ref{sec:physics}, we provide the analytical derivation of gap opening condition in a disc with MHD wind. Numerical methods and model parameters are described in Sect.~\ref{sec:methods}. We compare the results obtained with the 1D and 2D models in Sect.~\ref{sec:1d_vs_2d}. The results of a parameter space study conducted with the 1D code DEO are presented in Sect.~\ref{sec:param_study}. We then discuss the possible implication of the obtained results for the interpretation of the observational results in Sect~\ref{sec:discussion}. Finally, the conclusions of this work are summarised in Sect.~\ref{sec:conclusion}.

\section{Analytical estimates}\label{sec:physics}

\cite{LinPap86} have shown that the specific torque $\Lambda_{\rm p}$ from the planet on the surrounding gas can be adequately approximated by 
\begin{equation}
    \Lambda_{\rm p} = f \;\text{sign} \left(R-a\right)\; \frac{q^2}{2} \frac{a^4}{\Delta R^4} v_K^2\;,
    \label{Lambda0}
\end{equation}
where $v_{\rm K}$ is the Keplerian velocity at the radial distance of the planet $a$, $\Delta R = R-a$, and $0.1 \leq f < 1$ is a constant of order unity that depends on the detail of torque deposition and wave energy dissipation in the disc (e.g., \citealt{LinPap79}: $f\sim0.15$; \citealt{GoldreichTremaine80}: $f\sim0.4$; \citealt{RafikovPetrovich12}: $f\sim0.1$; \citealt{ArmitageNatar-02}; \citealt{DipierroLaibe17}). \bref{Based on numerical experiments that include planet migration (Wu et al, in prep.), in this paper we take $f=0.15$.}

To derive the conditions for a gap opening in a disc with MHD disc winds, we shall assume for simplicity that $\alpha_{\rm v}$ is negligibly small. 
At $R > a$, the angular momentum gain by the gas due to the disc-planet interaction results in an outward flow of gas at velocity
\begin{equation}
    v_{\rm p} = \frac{2\Lambda_{\rm p}}{v_K}\;.
\end{equation}
This outward flow of gas, away from the planet, results in the creation of the gap.
Due to 3D effects in realistic discs, the torque $\Lambda_{\rm p}$ from the planet saturates at distance $|\Delta R| \sim \kappa R_H$ away from the planet, where $\kappa$ is a constant of order unity. 

The MHD disc winds results in an inward flow of gas at the velocity \citep[][]{Tabone22-general}
\begin{equation}
        v_{\rm dw} = -\frac{3}{2}\alpha_{\rm dw} \left(\frac{H}{R}\right)^2 v_K\;.
    \label{v_dw}
\end{equation}
where $\alpha_{\rm dw}$ is a dimensionless parameter, defined by analogy with the $\alpha_{\rm v}$ parameter.
MHD winds will close a gap opened by the planet when $v_{\rm dw} + v_{\rm p} \leq 0$. Solving the equation  $v_{\rm dw} + v_{\rm p} =0$, we get the critical $q$ to open a gap in such a disc:
\begin{equation}
    q_{\rm dw} = \left[ \frac{\kappa^4}{2 f 3^{1/3}}\right]^{3/2} \alpha_{\rm dw}^{3/2} h^3\;,
    \label{qdw0}
\end{equation}
where $h=H/R$.
The result depends strongly on the parameter $\kappa$, which is not possible to constrain from first principles. We find that \bref{eq. \ref{qdw0} fits well the gap opening mass determined numerically (see the dashed line in Fig. \ref{fig:4} below) for $\kappa = 1.65$. With this value for $\kappa$,}
\begin{equation}
    q_{\rm dw} \approx 7 \times 10^{-5} \; 
    \left[\frac{ \alpha_{\rm dw}}{0.01}\right]^{3/2} 
     \left[\frac{ h}{0.1}\right]^{3} 
     \label{q_dw}
\end{equation}
This equation is interesting to compare with the gap opening mass in a viscous disc with $\alpha_{\rm dw}= 0$, eq.~(\ref{q_visc}). We can see that for $\alpha_{\rm v} = 10^{-3}$, $\alpha_{\rm dw} > 0.035$ is required for $q_{\rm dw}$ to exceed $q_{\rm visc}$. This predicts that MHD disc winds in a sense are less efficient in closing gaps opened in the discs by massive planets: for fiducial parameters used in eqs.~(\ref{q_visc}) and (\ref{q_dw}), $\alpha_{\rm dw} \gg \alpha_{\rm v}$ is required to close the gap from a given mass planet. 

Qualitatively, this result could be understood by looking at time scales for a gap of size $\Delta R$ to be closed. The turbulent angular momentum transfer is a diffusion process, and hence the gap closing time scales as $\propto \Delta R^2/\alpha_{\rm v}$. An MHD disc winds is an advective process, with the gap closing time scaling as $\Delta R/|v_{\rm dw}| \propto \Delta R/\alpha_{\rm dw}$. As $\Delta R \ll R$, that is, small, the turbulent gap closing timescale is much shorter than the MHD disc wind one at $\alpha_{\rm v} \sim \alpha_{\rm dw}$.

We define a modified parameter for gap opening that combines the previously derived \citep{CridaEtal06} criteria for gap opening in a turbulent disc with the condition we derived above for a disc with MHD disc wind parameter $\alpha_{\rm dw}$:
\begin{equation}
    \mathcal{P}_{\rm dw} =  \frac{3H}{4R_{\rm H}} + \frac{50 \alpha_{\rm v}}{q} \left({\frac{H}{R}}\right)^2  + \frac{70\alpha_{\rm dw}^{3/2}}{q} \left({\frac{H}{R}}\right)^3\;.
\label{Crida_mod}
\end{equation}

\section{Numerical methodology}\label{sec:methods}

In this paper, we use both one-dimensional (1D) and two-dimensional (2D) simulations to study gap opening by planets in discs with MHD disc winds. The latter simulations are more rigorous and yield more accurate results than the former, but are relatively numerically expensive. The 1D simulation method is less rigorous but is fast, permitting quick surveys of the large parameter space of this problem. We shall see below that the 1D method does capture the main traits of the planet gap opening in the presence of MHD disc winds.

\subsection{1D code DEO}\label{sec:1d}

The master equation for discs with angular momentum transfer due to both turbulent viscosity and MHD winds \citep[e.g.,][]{Tabone22-general}, and the planet tidal torques, in 1D is 
\begin{equation}
\begin{split}
    \frac{\partial\Sigma}{\partial t} = \frac{3}{R} \frac{\partial}{\partial R} \left[ R^{1/2} \frac{\partial}{\partial R} \left(R^{1/2}\nu \Sigma\right) \right] 
    -\frac{1}{R} \frac{\partial}{\partial R} \left( v_{\mathrm{dw} }\Sigma R \right)  - \\ - \frac{1}{R} \frac{\partial}{\partial R} \left(2\Omega^{-1} \Lambda_{\rm p} \Sigma\right)\;,
\end{split}
\label{dSigma_dt}
\end{equation}
where $\Sigma$ is the gas surface density, $\Omega$ is the angular velocity at radial distance $R$, $\nu=\alpha_{\rm v} c_{\rm{s}} H$ is the kinematic viscosity of the disk, $c_{\rm{s}}$ is the gas sound speed at the disk midplane, and $H$ is the vertical scale height of the disk. As many authors in the past \citep[e.g.,][]{Armitage13-MHD-wind,Hasegawa17-MHD-wind}, we neglected the mass loss term from the disc due to the wind in eq.~(\ref{dSigma_dt}). This is rigorously correct only in the limit of the lever-arm parameter of the wind, $\lambda_{\rm w} \rightarrow \infty$ \citep[cf. the last term in eq. 10 of][]{Tabone22-general}. However, here we focus on the depth of the gap open by the planet in its vicinity, in the region $|R-a| \sim R_{\rm H}$. The radial distance of the planet is fixed at $a=45$~AU; we note that this value is arbitrary and has no implications for our conclusions. The ratio of the mass loss term to the second term on the right-hand side of eq.~(\ref{dSigma_dt}) is $\sim R_{\rm H}/[2R(\lambda_{\rm w} -1)]$. This is much smaller than unity for $\lambda_{\rm w}-1 \gg R_{\rm H}/2R \sim q^{1/3}$. Considering that from energetic arguments $\lambda_{\rm w}>3/2$ \citep[e.g.,][]{1982BlandfordPayne}, it appears that this condition should always be satisfied for physically motivated values of $\lambda_{\rm w}$ and that omission of the wind loss term is well justified (in the same way as, for example, photo-evaporation is not usually taken into account when studying gap-opening). \bref{Due to its relative unimportance, the inclusion of the mass loss term to eq. \ref{dSigma_dt} would not modify the main conclusions of our paper, and would only expand the parameter space of the problem unnecessarily. For these reasons,} we leave the consideration of the mass-loss term influence on the gap opening mass to future work.



We use the "evanescent" boundary conditions for the disc, keeping the surface densities of both gas and dust constant at the outer boundary of our computational domain (for both 1D and 2D simulations), and we  neglect planet accretion or migration. These simplifying approaches are widely used in the field \citep[e.g., see][]{RosottiEtal16,2018ZhangDSHARP,Liu18-HD163296}.

The 1D disc equations are solved with the code DEO (Disc with an Embedded Object). The code is described in a recent paper by \cite{2022NayakshinElbakyan}, and we only briefly summarise its main features.  We use the well known \citep[e.g.,][]{LinPap86} prescription for the tidal torque from the planet on the disc.  We solve dust dynamics for fixed Stokes number dust particles, assuming that the dust radial velocity is given by the terminal velocity approximation plus the dust diffusion term due to disc turbulence with the turbulent viscosity parameter $\alpha_{\rm v}$. This also presumes that dust back reaction force on the gas is negligible.

We assume that the disc temperature profile is given by 
\begin{equation}
     T = T_0 \left(\frac{R_0}{R}\right)^{1/2}
    \label{T-profile}
\end{equation}
where $R_0 = 100$~AU and $T_0 = 10$~K.
With the mass of the central star set to $M_{\star} = 1~\msun$, the disc vertical aspect ratio is $H/R = h_0 (R/R_0)^{1/4}$, where $h_0 = (k_b T_0 R_0/GM_*\mu)^{1/2} = 0.08$, $k_b$ is the Boltzmann's constant and $\mu = 2.4m_p$ is the mean molecular weight.

The initial gas surface density profile at time $t=0$ is given by
\begin{equation}
    \Sigma(R) = \Sigma_0 \left(\frac{R_0}{R}\right)^{-1}\;,
    \label{sigma-disc0}
\end{equation}
where $\Sigma_0$ is a normalisation constant. With our choice of flaring index, constant $\alpha_{\rm dw}$ and $\alpha_{\rm v}$, and the neglect of the mass-loss term (i.e., $\lambda_{\rm w}\rightarrow \infty)$, this is a steady state solution of eq.~(\ref{dSigma_dt}) without the planet.
The initial surface density of dust is given by
\begin{equation}
    \Sigma_{\rm d}(R) = \epsilon \Sigma\;,
    \label{sigma-dust}
\end{equation}
where $\epsilon = 0.01$ is the dust-to-gas mass ratio. The evanescent boundary conditions enforce constant $\Sigma(R_{\rm out})$ and $\Sigma_{\rm d}(R_{\rm out})$ at the outer boundary of the computational domain (200 AU). \bref{The 1D simulations are terminated when the age of the system reaches 300 kyr.}

\begin{table}
\centering
\caption{\label{tab:1} Basic parameters of the models.}
\begin{tabular}{p{0.15\linewidth} p{0.25\linewidth} p{0.45\linewidth}}
\hline 
\hline 

Parameter & Value & Description \tabularnewline
\hline 
$a$ & 45 AU & Radial distance of the planet \tabularnewline
$R_0$ & 100 AU & Radial distance at which the normalisation constants are calculated \tabularnewline
$T_0$ & 10 K & Disc temperature at the radial distance $R_0$ \tabularnewline
$\Sigma_0$ & 7.2$\times10^{-2}$~g~cm$^{-2}$ & Gas surface density at the radial distance $R_0$
\tabularnewline
$h_0$ & 0.08 & Disc aspect ratio at the radial distance $R_0$ \tabularnewline
$\epsilon$ & 0.01 & Dust-to-gas mass ratio in the disc \tabularnewline
$R_{\rm in, DEO}$ & 1 AU & Inner boundary of the computational domain in the 1D model \tabularnewline
$R_{\rm out, DEO}$ & 225 AU & Outer boundary of the computational domain in the 1D model \tabularnewline
$R_{\rm in, FARGO}$ & 13.5 AU & Inner boundary of the computational domain in the 2D model \tabularnewline
$R_{\rm out, FARGO}$ & 225 AU & Outer boundary of the computational domain in the 2D model \tabularnewline
$M_*$ & 1 $\msun$ & Mass of the central star \tabularnewline
$t_{\rm sim}$ & 300 kyr (1000 orbits) & Duration of the simulations  \tabularnewline
\hline 
\end{tabular}
\end{table}

\subsection{2D simulations}\label{sec:2d}

To perform two-dimensional simulations of planets embedded in discs with MHD disc winds, we use the well known public code FARGO3D \citep{BL_Masset_16_FARGO3D}. Dust grains are modelled as pressureless fluid \citep[see][]{BL19-FARGO}. We employ 512 logarithmically spaced zones and 656 uniformly spaced grids zones in the radial and azimuthal directions, respectively, with an inner and outer radii of $0.3 a$ and $5 a$, respectively.
We activated the STOCKHOLM option in FARGO3D in order to minimise the artefacts of spurious waves reflection on the boundaries of the computational domain. The initial disc gas and dust surface density profiles, and the disc temperature are the same as for our 1D model. \bref{The 2D simulation runs for 1000 orbits at the position of the planet, which is 300 kyr.
Main initial parameters used in our models are presented in Table~\ref{tab:1}.}

The public version of FARGO3D does not include an option for an MHD disc wind torques. In our code, we extend the equations of \cite{Tabone22-general} from 1D to 2D, and impose a specific torque $\Gamma$ on the gas,
\begin{equation}
    \Gamma = \frac{1}{2} \sqrt{ \frac{GM_*}{R}} v_{\rm dw}\;,
    \label{Gamma0}
\end{equation}
where $v_{\rm dw}$ is given by eq.~(\ref{v_dw}). As with the 1D treatment, we neglect the mass loss by the disc.

\section{Comparison of gaps in 1D and 2D} \label{sec:1d_vs_2d}


\begin{figure*}
\begin{centering}
\includegraphics[width=2\columnwidth]{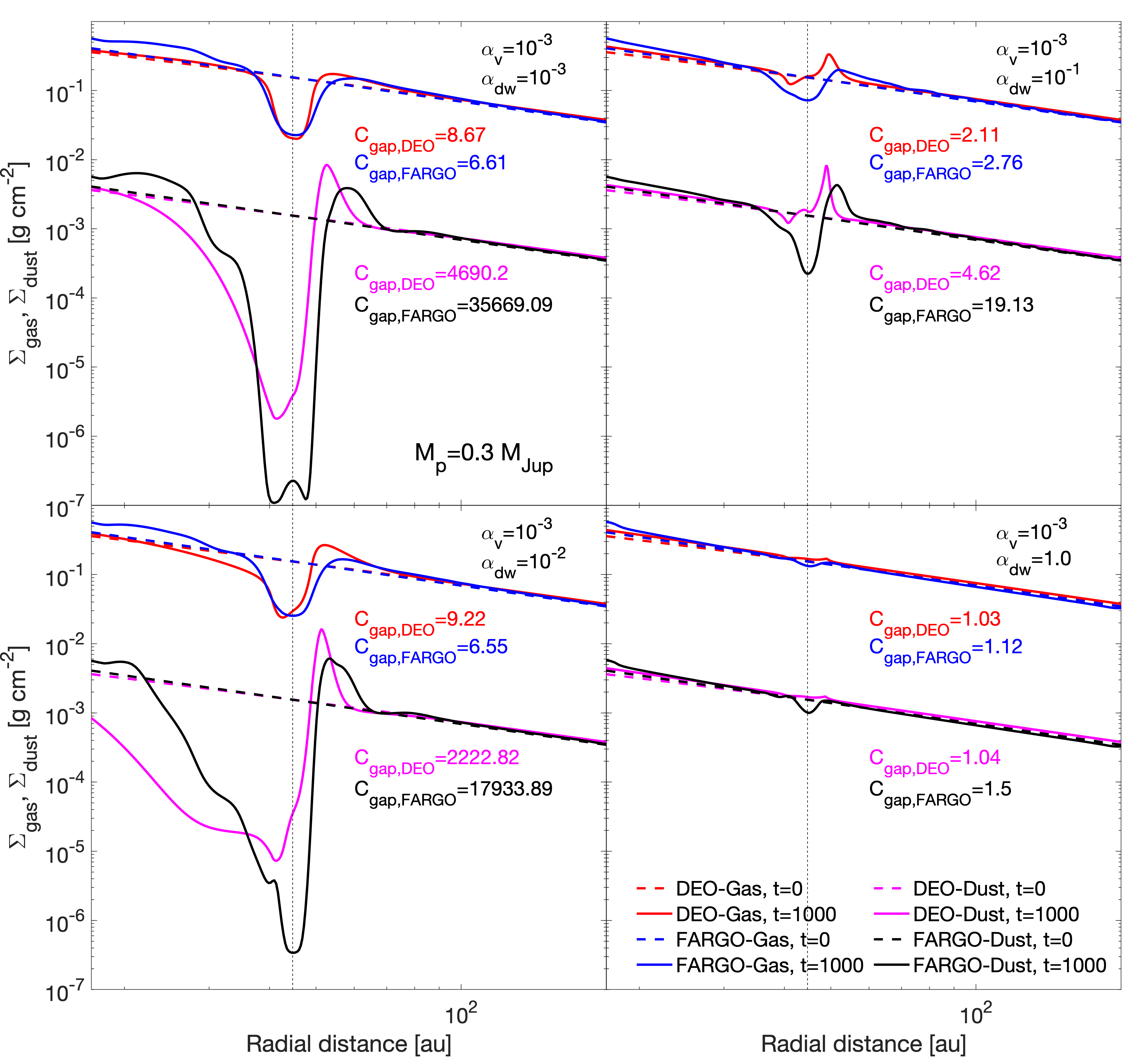}
\par\end{centering}
\caption{\label{fig:1} Gas and dust surface density profiles calculated with the 1D code DEO and 2D code FARGO. The profiles for 2D model are calculated by azimuthally averaging the 2D data. Each panel represents models with a distinct value of $\alpha_{\rm dw}$, shown in the top right corner. We parametrise the turbulent viscosity in all models with $\alpha_{\rm v}=10^{-3}$. The dashed lines show the initial surface density distribution, while the solid lines show the distribution after 1000 orbits at radial distance of the planet, which is marked with the vertical dashed line. \bref{All dust particles have Stokes number $\St = 10^{-2}$.}}
\end{figure*}



The depth of the gap opened by the planet can be measured by the ratio of the perturbed density profile at the planet location to the unperturbed one \citep[e.g.,][]{CridaEtal06}. However, the gas surface density at the location of the planet may evolve not only due to the planet itself but also due to other factors such as the MHD disc wind, or the boundary conditions, etc. \bref{In some cases, the surface density of dust can show a buildup at the orbit of the planet, instead of a gap, when the strong MHD winds are present in the disc. This buildup is a result of dust accumulation outside the planet orbit. In this case, no gap is present in the dusty disc, following the gap definition by \citet{CridaEtal06}. However, the dusty disc is strongly perturbed by the presence of the planet, and to take into account the impact of planet onto the disc, we use slightly different definition for the gap. } In this situation a better quantity to use in characterising the gap is gap contrast defined by 
\begin{equation}
    C_{\mathrm{gap}} = \frac{\Sigma_{\mathrm{max}}}{\Sigma_{\mathrm{min}}}.
    \label{eq:Cgap}
\end{equation}
where $\Sigma_{\rm max}$ is the peak value of the gas (dust) surface density outside the planet orbit and $\Sigma_{\rm min}$ is the minimum value of the gas (dust) surface density inside the gap. Below, we define the gap contrast separately for 1D ($C_{\rm gap,DEO}$) and 2D ($C_{\rm gap,FARGO}$) runs.


Fig.~\ref{fig:1} shows the gas and dust surface density profiles in the 1D and 2D discs. The profiles for 2D models are calculated by azimuthally averaging the surface density value. 
We exclude the matter inside the Hill sphere of the planet during the azimuthal averaging. 
For both 1D and 2D models, we use turbulent viscosity parameter $\alpha_{\rm v}=10^{-3}$. Models with different strength of turbulence in the disc are discussed in Sect.~\ref{sec:param_study}. Each panel of the figure represents a model with a distinct $\alpha_{\rm dw}$ value presented in the top right corner of the panel. The dashed lines show the initial gas and dust surface density profiles in the models, while the solid lines show the surface density profiles after 1000 orbits at the position of the planet. Surface density of gas is shown with the red and blue lines for the 1D and 2D models, respectively, whereas the dust surface density is plotted with the magenta and black lines. The vertical dashed line marks the position of the planet. 

The dusty component in our models is represented with the dust particles that have fixed Stokes number $\rm{St}=10^{-2}$. This value is widely used as threshold value for the pebble definition \citep{2012LambrechtsJohansen, 2019LenzKlahr}. Mass accumulation of dust particles at the outer edge of the gap is present in both 1D and 2D models. The pressure bump formed at the edge of the planetary gap is a possible location for trapping the inward migrating pebbles \citep{2020Eriksson,2021Carrera}, which play a crucial role in the mass accumulation of the planet by the pebble accretion \citep[e.g.,][]{2017JohansenLambrechts}. We note that in our models the planet mass is $M_{\rm p}=0.3~M_{\rm Jup}$ and stays constant.

The values of gap contrast $C_{\rm gap,DEO}$ and $C_{\rm gap,FARGO}$ for gas (red and blue) and dust (magenta and black) are presented in each panel of Fig.~\ref{fig:1}. We define that the gap is opened in the gas (dusty) disc when the $C_{\rm gap} = \Sigma_{\rm max}/\Sigma_{\rm min}=10$. As explained above, this definition is slightly different from the one used in \cite{CridaEtal06}, however we find no significant difference between the models with different gap definition. \bref{Quantitively, the value of $C_{\rm gap}$ differs by 5-70\% between the different gap definitions, however, this does not impact the main results of current study. The comparison between the two different gap definitions is presented in Sect.~\ref{sec:param_study}.}


The planet opens a partial gap in the gas for both 1D and 2D models shown in Fig.~\ref{fig:1}. The local depression around the planet position is considered as a partially opened gap, since the gap contrast stays less than the threshold value for the gap opening $C_{\rm gap}=10$. Gap contrasts for the gas disc in the 1D and 2D models differ by less than a factor of 1.4, which means that the 1D code reproduces the gap opening process quite precisely. Moreover, the shapes of the partially opened gas gaps in the 1D and 2D models look quite similar for the discs with $\alpha_{dw}=10^{-3}$ and $\alpha_{dw}=10^{-2}$. The shapes of the gas gaps in the 1D and 2D models do differ for the discs with higher $\alpha_{dw}$ values. Nevertheless, we note that the width of the gap and the gap contrast are similar in our 1D and 2D simulations.


In contrast to the gas gaps, the gaps in the dusty disc are deep with $\alpha_{\rm dw}=10^{-3}$ and $\alpha_{\rm dw}=10^{-2}$ for both 1D and 2D models. The values of $C_{\rm gap}$ in the dust are a few orders of magnitude higher than the threshold value for gap opening, which means that practically no dust is left inside the gap. On the other hand, dust particles are accumulated at the outer edge of the gap both in 1D and 2D models. In the 1D model with $\alpha_{\rm dw}=10^{-1}$, the planet opens a partial gap in dust. At the same time, in the 2D model with $\alpha_{\rm dw}=10^{-1}$, the gap is opened fully. However, the gap contrasts in 1D and 2D models differ less than by a factor of 4. The planet does not open a gap in the models with $\alpha_{dw}=1.0$ and only creates a moderate local depression in the dust surface density.

\begin{figure}
\begin{centering}
\includegraphics[width=1\columnwidth]{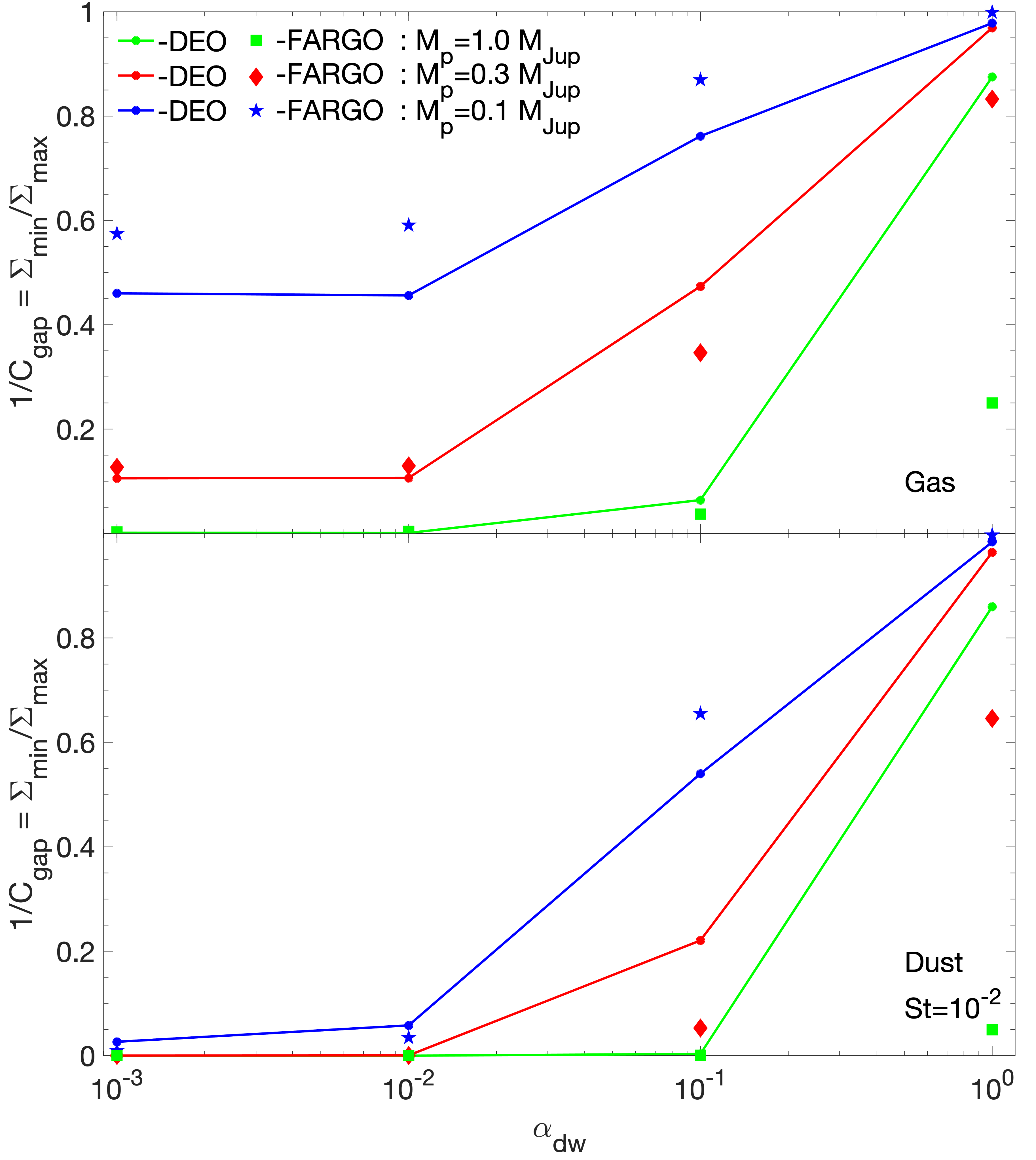}
\par\end{centering}
\caption{\label{fig:2} Dependence of 1/$C_{\rm gap}$ on $\alpha_{\rm dw}$ values for the gas (top panel) and the dust (bottom panel) component. Solid lines show the results obtained by the 1D code DEO, while the independent markers show the results from the 2D code FARGO. Each colour corresponds to a distinct planet mass.}
\end{figure}

In addition to the model with the 0.3~$M_{\rm Jup}$ mass planet, we calculate the evolution of the disc with a higher mass planet (1.0~$M_{\rm Jup}$) and a lower mass planet (0.1~$M_{\rm Jup}$). The resulting gap properties at 1000 planet orbits are shown in Fig.~\ref{fig:2}  for all three planet masses. Here we plot the ratio $\Sigma_{\rm min}/\Sigma_{\rm gap} = C_{\rm gap}^{-1}$ vs $\alpha_{\rm dw}$ on a linear scale. The results for 1D models are shown with the solid lines, while the results for 2D models with symbols. The lines and markers with green, red and blue colours are representing results for the planets with 1.0, 0.3, and 0.1~$M_{\rm Jup}$, respectively.


The agreement between the 1D and 2D models is within a factor of a few for all the values of $\alpha_{\rm dw}$, except for the highest $\alpha_{\rm dw}=1.0$ case, which we shall argue is likely physically irrelevant in any case. 


\section{Parameter space study} \label{sec:param_study}

\subsection{Gaps in the gas disc}\label{sec:gas_gap}


While comparing 1D and 2D models in Sect.~\ref{sec:1d_vs_2d}, we calculated gap contrasts for planets of three different masses and four different $\alpha_{\rm dw}$ values. We now sample a broader parameter space with much finer parameter spacing, using 50 planet masses varying from $10^{-2}$ to 90 $M_{\rm Jup}$ and with 50 $\alpha_{\rm dw}$ values in a range of between $10^{-4}$ and 1.0. Thus, the $\alpha_{\rm dw}$ - $M_{\rm p}$ phase space is covered with 2500 disc models calculated with the 1D code. It is currently prohibitively expensive to calculate such a number of models in 2D with FARGO3D.


\begin{figure}
\begin{centering}
\includegraphics[width=1\columnwidth]{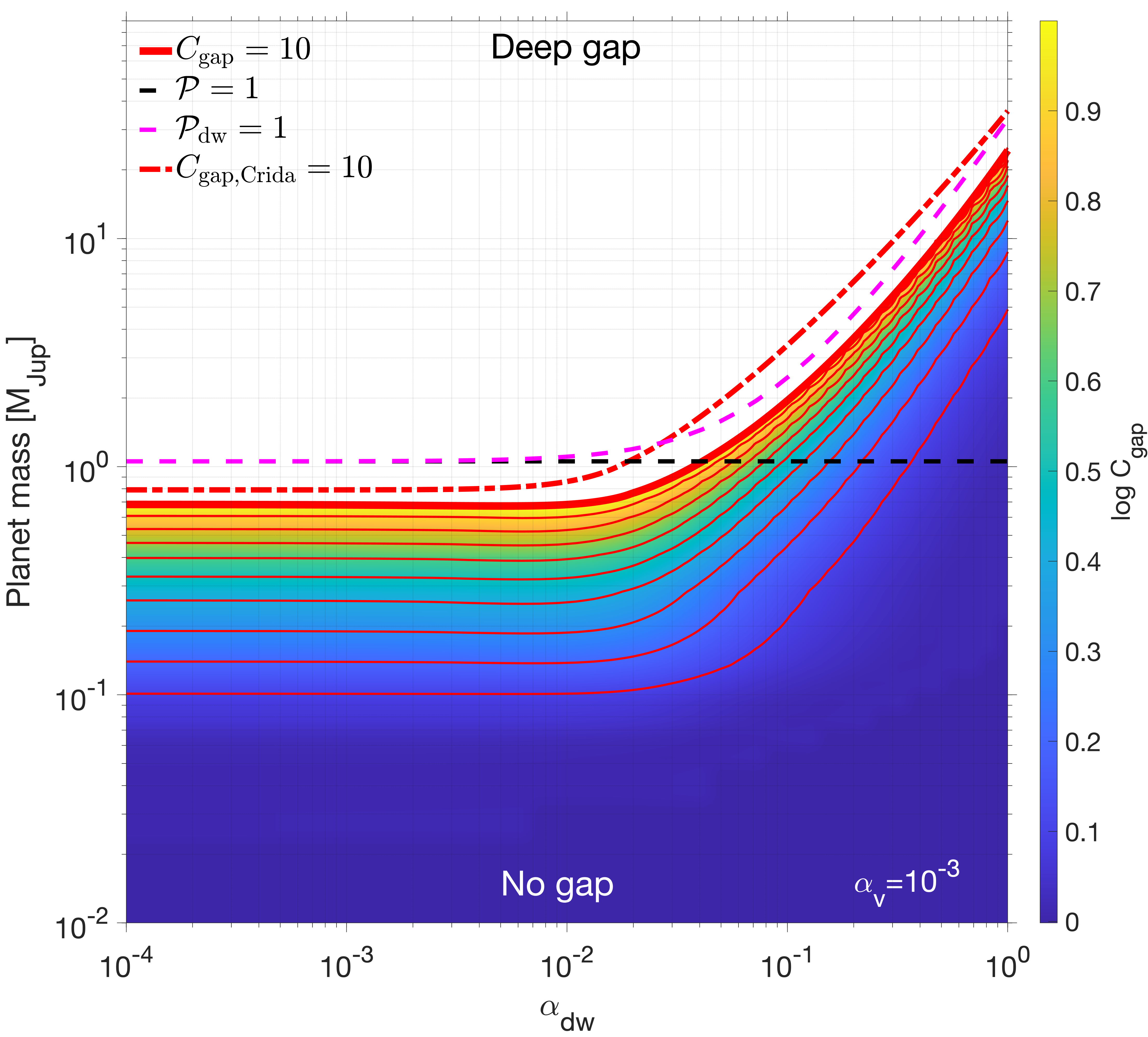}
\par\end{centering}
\caption{\label{fig:3} The gap contrast $C_{\rm gap}$ for the gas disc with the $\alpha_{\rm v}=10^{-3}$ shown on the $\alpha_{\rm dw}$ - $M_{\rm p}$ phase space with the colour. The red contours show log($C_{\rm gap}$) from 0.1 to 1.0 (bottom to top) with a 0.1 step. The thick red line shows the $C_{\rm gap}=10$ values, which correspond to the planet masses, $M_{\rm gap}$, needed to open a gap in the gas disc. \bref{The dash-dotted red line shows the $C_{\rm gap}=10$ values, but calculated using the gap definition from \citet{CridaEtal06}.} The dashed black and magenta lines show the analytical $M_{\rm gap}$ values for the gas disc calculated for the $\mathcal{P}=1$ (eq.~(\ref{CridaP0})) and $\mathcal{P}_{\rm dw}=1$ (eq.~(\ref{Crida_mod})), respectively. The $C_{\rm gap}>10$ values are not shown on the figure for a clarity.}
\end{figure}

Fig.~\ref{fig:3} shows the gap contrast $C_{\rm gap}$ with a colourmap in the $M_{\rm p}-\alpha_{\rm dw}$ parameter space for the turbulent viscosity parameter $\alpha_{\rm v}=10^{-3}$. The contour lines show log($C_{\rm gap}$) from 0.1 to 1.0 (bottom to top) with a step equal to 0.1. The contour line for log($C_{\rm gap}$)=1, that is, $C_{\rm gap} =10$, is shown with the thick red line; this line shows the corresponding $M_{\rm gap}$. In the white region of the colourmap, the gap is considered fully opened, whereas in the dark regions the gap is barely visible.


The dashed black and magenta lines show $M_{\rm gap}$ values for which $\mathcal{P}=1$ or $\mathcal{P}_{\rm dw}=1$, respectively. The former is appropriate for discs without MHD disc winds ($\alpha_{\rm dw}=0$). Note that $M_{\rm gap}$ can be found in a closed analytical form by solving eq. (\ref{Crida_mod}) for $\mathcal{P}_{\rm dw}=1$,
\begin{equation}
    q = \frac{M_{\mathrm{gap}}}{M_*} = 2 Y[(X+1)^{1/3}-(X-1)^{1/3}]^{-3},
    \label{qmin_anlytic}
\end{equation}
with $X=\sqrt{1+3h^3/(16Y)}$ and $Y=50\alpha_{v}h^2+70\alpha_{\rm dw}^{3/2}h^3$. The dashed lines overestimate $M_{\rm gap}$ (the solid red curve) by $\sim $20-50\% over the whole parameter space studied in Fig.~\ref{fig:3}. \bref{The magnitude of this deviation is reasonably small given the following. Firstly, the gap opening criteria from \citet{CridaEtal06} is a linear fit to their numerical data which itself has an accuracy of order $\sim 5-30$\% (see fig. 12 in their paper). Secondly, their study is in 2D while in Fig. 3 we show results of our 1D code. Finally, our gap opening criterion is slightly different from the one used by \citet{CridaEtal06}. Overall, we consider the agreement between the solid red and the dashed magenta curves encouraging; in the limit $\alpha_{\rm dw} \rightarrow 0$ our 1D disc models are able to reproduce the gap opening mass calculated based on 2D results of \cite{CridaEtal06}  reasonably well.}



We observe that $M_{\rm gap}$ is independent of $\alpha_{\rm dw}$ for $\alpha_{\rm dw}\lesssim 10^{-2}$, that is, the role of disc wind in closing the gap is negligible for small  $\alpha_{\rm dw}$. This fact illustrates that the residual disc turbulence may be key in closing the gap unless $\alpha_{\rm dw}$ is sufficiently high.

\begin{figure}
\begin{centering}
\includegraphics[width=1\columnwidth]{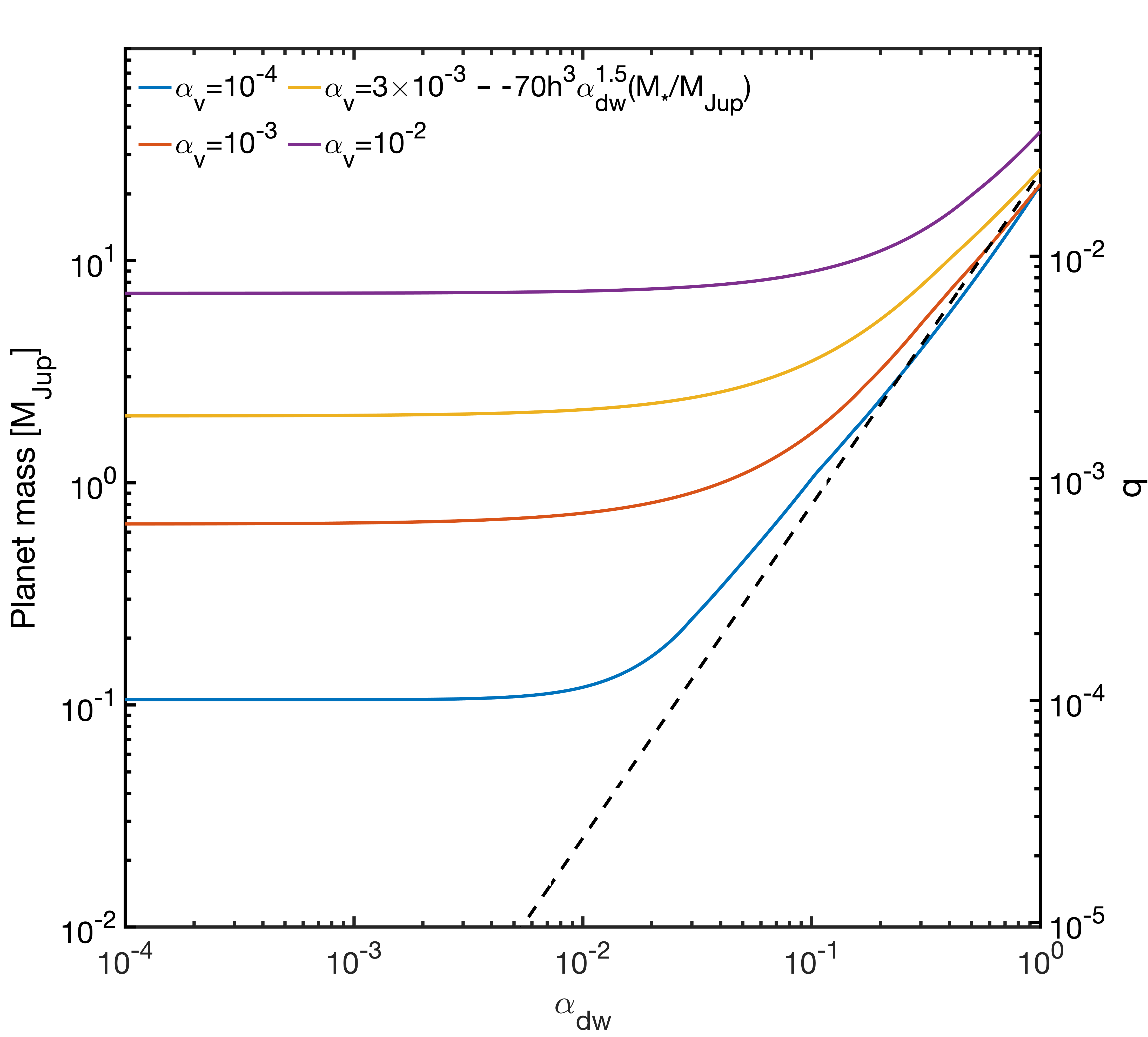}
\par\end{centering}
\caption{\label{fig:4} The dependence of minimum planet mass needed to open a gap in the gas disc, $M_{\rm gap}$, on the $\alpha_{\rm dw}$ parameter. Each solid line corresponds to a distinct value of $\alpha_{\rm v}$ in the disc. The dashed line shows the analytical power law dependence of planet mass on $\alpha_{\rm dw}$ (last term on the rhs of Eq.~(\ref{Crida_mod})).}
\end{figure}

We now consider how these results depend on the value of $\alpha_{\rm v}$. To this end, we repeat the parameter space calculations done for Fig.~\ref{fig:3}, but now for several additional values of $\alpha_{\rm v}=10^{-4}$, $3\times10^{-3}$, and $10^{-2}$.  Fig.~\ref{fig:4} presents corresponding $M_{\rm gap}$ vs $\alpha_{\rm dw}$ with the coloured solid lines. Note that the  red  $\alpha_{\rm v}=10^{-3}$ curve is the same as the thick red curve in Fig. \ref{fig:3}. The dashed line in Fig.~\ref{fig:4} shows eq.~(\ref{q_dw}), i.e., $M_{\rm gap} = q_{\rm dw} \msun$, which is the limit in which the disc wind completely dominates the gap closing process.

Taken together, Figs. \ref{fig:3} and \ref{fig:4} show that MHD disc winds play a negligible role in the gap closing process unless $\alpha_{\rm dw} \gtrsim 10^{-2}$. If the value of the turbulent viscosity parameter is as small as ALMA observations suggest (see references in Introduction) then gaps opened by a Jupiter mass planet can be closed only if the MHD disc wind parameter $\alpha_{\rm dw}$ exceeds $\sim 0.1$, which we argue (cf. Sect.~\ref{sec:discussion}) is unlikely in the steady state for most of the observed discs. 

\begin{figure}
\begin{centering}
\includegraphics[width=1\columnwidth]{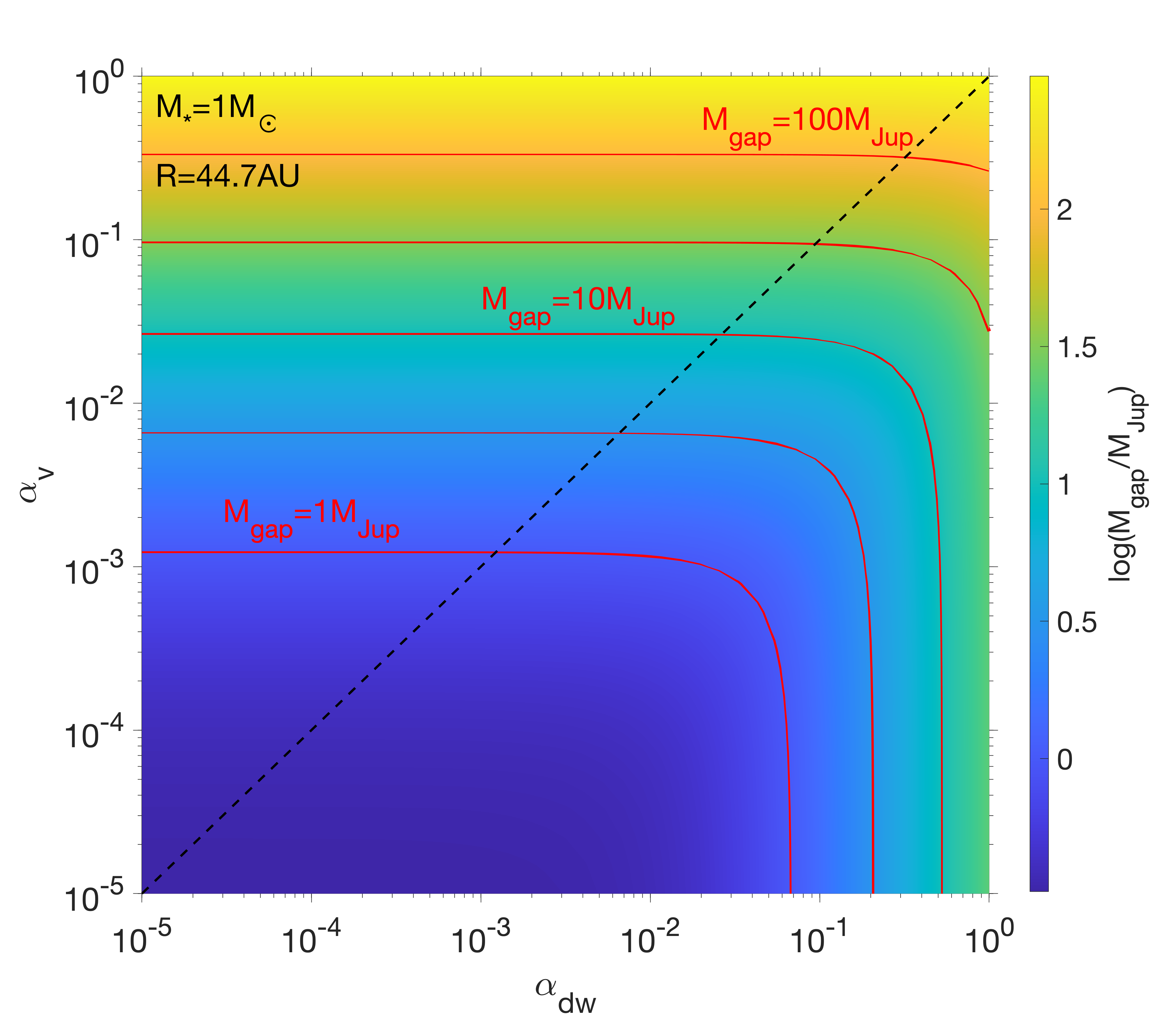}
\par\end{centering}
\caption{\label{fig:4a} The gap opening planet mass, $M_{\rm gap}$, shown with the colour on the $\alpha_{\rm dw}$-$\alpha_{\rm v}$ parameter space. The planet orbits around a $M_*=1\msun$ star at the radial distance $R=45$~AU. The red contours show the log($M_{\rm gap}$) values from 0 to 2 with 0.5 step. The dashed line corresponds to the $\alpha_{\rm dw}$=$\alpha_{\rm v}$ values.}
\end{figure}

Finally, Fig. \ref{fig:4a} shows $M_{\rm gap}$ with a colourmap calculated with eq.~(\ref{qmin_anlytic}) for $M_*=1 \msun$ and the chosen disc model with the planet at $R = a$=45~AU (i.e., the radial distance of the planet in Sect.~\ref{sec:1d_vs_2d}) for a broad range of $\alpha_{\rm dw}$ and $\alpha_{\rm v}$. The disc aspect ratio at the position of the planet is $h=0.05$. The contour lines represent the planet isomass contours, i.e., the lines on which $M_{\rm gap} =$ const, and correspond to log($M_{\rm gap}$) values from 0 to 2 with a step of 0.5. The dashed curve shows the line $\alpha_{\rm dw}$=$\alpha_{\rm v}$.  The isomass contours yet again indicate that discs winds are inefficient in closing the gaps compared with turbulence. For example, consider the lowest isomass contour, where $M_{\rm p}=1 \mj$. We can see that \bref{such a planet is unable to open a gap}
for $\alpha_{\rm v} \geq 10^{-3}$. However, if disc viscosity is much smaller than that, then the same planet manages to maintain a deep gap in the disc for  $\alpha_{\rm dw}$ values as large as $\alpha_{\rm dw} \leq 5\times 10^{-2}$.

\subsection{The gap opened in dust disc}\label{sec:dust_gap}

\begin{figure}
\begin{centering}
\includegraphics[width=1\columnwidth]{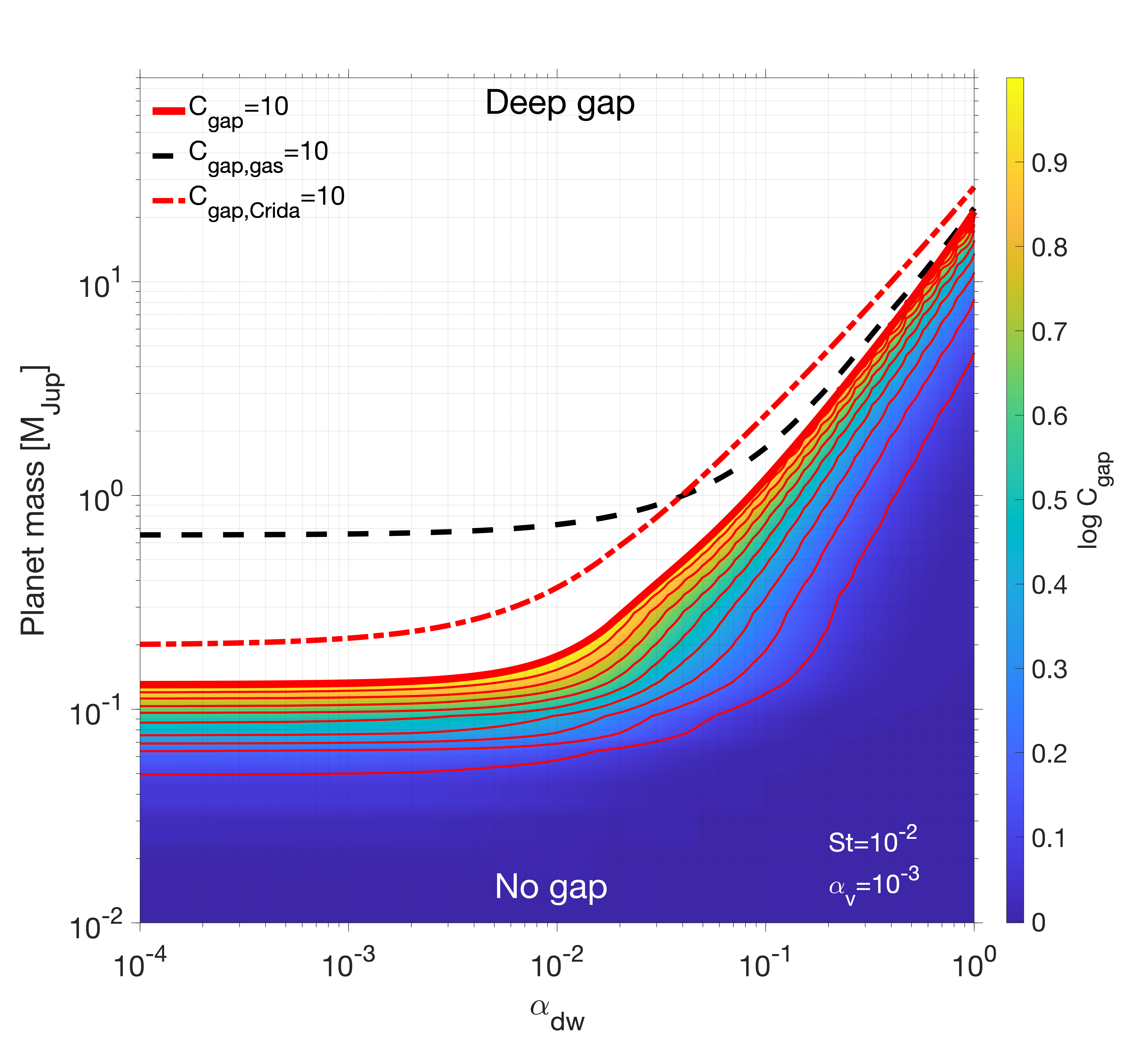}
\par\end{centering}
\caption{\label{fig:5} Similar to Fig.~\ref{fig:3}, but for the dusty disc with the fixed Stokes number $\rm{St}=10^{-2}$ for the dust particles. The dashed black line is identical to the thick red line from Fig.~\ref{fig:3}, which shows the planet mass needed to open a gap in the gas disc.}
\end{figure}

Disc substructures, such as rings and gaps, are usually observed through the thermal emission of mm-sized dust particles \citep[e.g.,][]{2018LongPinilla, 2018HuangAndrews}. Here we present a preliminary study of the gap opening process in the dust.  Fig.~\ref{fig:5} shows the gap contrast $C_{\rm gap}$ calculated with the 1D approach for the dust component with a fixed Stokes number, $\rm{St}=10^{-2}$. As in Fig.~\ref{fig:3}, we set the viscosity parameter at $\alpha_{\rm v} = 10^{-3}$. The dashed black line in Fig.~\ref{fig:5} is the same as  the thick red line in Fig.~\ref{fig:3}. This line shows the gas disc $M_{\rm gap}$ for reference. 

Two major conclusions follow from comparison of the red and the dashed black curves in Fig. \ref{fig:5}. When the disc wind is weak, $\alpha_{\rm dw}\lesssim10^{-2}$, the planet mass needed to open a deep gap in the dust is a factor of 5 lower than that for the gas ($M_{\rm gap}$). However, for strong disc winds,  $\alpha_{\rm dw}> 10^{-2}$, the dust gap opening mass increases rapidly, and for $\alpha_{\rm dw} > 10^{-1}$ the gas and dust gap opening masses converge to approximately the same function.



This can be understood as the following. The radial velocity of dust in a steady state disc can be presented as \citep[e.g.,][]{2016Birnstiel}
\begin{equation}
    \label{eq:ur}
    u_{\rm r}\simeq u_{\rm drift}+u_{\rm adv},
\end{equation}
where the drift velocity
\begin{equation}
    u_{\rm drift}=-\frac{\eta h^2 \, \mathrm{St} \, v_{\rm K}}{1+\mathrm{St}^{2}}, \; \; \eta=-\frac{\partial \mathrm{ln} P}{\partial \mathrm{ln} R}\;,
    \label{udrift0}
\end{equation}
and $u_{\rm adv}$ is the advection velocity, the one with which the dust is carried in by the radial flow of gas  \citep{2002TakeuchiLin},
\begin{equation}
    u_{\rm adv}=\frac{v_{\rm r}}{1+\mathrm{St}^2}\;.
    \label{uadv0}
\end{equation}
Here $P$ is the disc midplane pressure, and $v_{\rm r}$ is the gas radial velocity. The radial drift of dust particles is due to the difference in the azimuthal velocities of gas and dust, forcing the dust to drift in the direction of increasing pressure \citep[e.g.,][]{Weiden77}. The gas radial velocity in a disc with both viscous and MHD disc wind angular momentum transfer is
\begin{equation}
        v_{\rm r} = -\frac{3}{2}\left( \alpha_{\rm v} + \alpha_{\rm dw}\right) h^2 v_K\;.
    \label{v_gas}
\end{equation}
Taking the ratio of the two contributions to the dust radial velocity,
\begin{equation}
   \left| \frac{u_{\rm drift}}{u_{\rm adv}}\right| = \frac{2\eta \St}{3(\alpha_{\rm v} + \alpha_{\rm dw})}\;.
\end{equation}
In a disc with $\eta \sim$ a few, and weak turbulence and MHD disc winds, dust particles with sufficiently large Stokes numbers always drift through the gas faster than $v_{\rm adv}$. However, for a strong MHD wind, and neglecting the disc turbulence ($\alpha_{\rm dw}\gg \alpha_{\rm v}$), the advective velocity becomes larger than the dust drift velocity if
\begin{equation}
     \alpha_{\mathrm{dw}}>\frac{2}{3}\eta \St\;.
\end{equation}
Additionally, since gas pressure in the disc is isotropic, the maximum radial pressure gradient should be of order $\sim P/H$, and hence $\max[ \partial \ln P/\partial \ln R] = 1/h$. Therefore, a sufficient condition for the dust to be mainly carried with the gas (rather than drift through it) is
\begin{equation}
     \alpha_{\mathrm{dw}} \geq \frac{2 \St}{3 h}\;.
     \label{alpha_drift}
\end{equation}
For our disc model, this condition yields $\alpha_{\rm dw} \gtrsim 0.1$, in good agreement with Fig. \ref{fig:5}.

Summarising, these results were derived for low disc turbulence, i.e., $\alpha_{\rm v} = 10^{-3}$. In this case, at weak disc winds, the planet mass necessary to open a deep gap in the dust disc is significantly lower than the gas $M_{\rm gap}$ value. At high $\alpha_{\rm dw}$ as given by eq.~(\ref{alpha_drift}), the dust is advected with the gas, and the gap opening mass is the same for gas and dust discs.
We caution the reader that these results for dust gaps are preliminary. A larger parameter space study performed in 2D is needed to confirm these 1D results. 

\bref{Furthermore, our main conclusion is that planet gap opening is more efficient in MHD disc wind dominated discs compared with standard viscous discs, \bref{if the discs have similar mass and angular transport rates}. This conclusion is based on the 1D code results, and the fact that gaps are somewhat deeper in 2D implies that this conclusion would only be made stronger if we were able to use 2D methods for a wide parameter study below. We therefore conclude that for practical reasons the 1D treatment is sufficiently accurate in estimating the depth of a gap opened by the planet.}

\section{Discussion}\label{sec:discussion}

In this paper, we have shown that tidal torques from planets open deep gaps in gas and dust discs much more readily if angular momentum transfer is dominated by MHD disc winds rather than turbulent viscosity. \bref{Consider a specific example of a disc with surface density profile $\Sigma$ at some radius $R$. Compare a disc powered by MHD disc winds with  $\alpha_{\rm dw}=10^{-3}$, and a negligible turbulent viscosity, $\alpha_{\rm v} = 0$, with the opposite limiting case of a turbulent disc with $\alpha_{\rm v}=10^{-3}$ and $\alpha_{\rm dw} =0$. A planet of same mass is much more likely to open a gap in the former rather than the latter even though the mass and angular momentum in both discs is transferred with the same rate.} 
Physically, this is because at the same radial flow velocity (that is, $\alpha_{\rm v} = \alpha_{\rm dw}$), advection of material across the gap region is a far less efficient mechanism of closing the gap than turbulent diffusion. This holds for both gas and dust (cf. Sect.~\ref{sec:dust_gap}). We derived a gap opening criterion which uses a dimensionless parameter $\mathcal{P}_{\rm dw}$ that takes into account both viscous and magnetised disc wind dominated mass transfer. For $\mathcal{P}_{\rm dw} \lesssim 1$ the gap is opened, and for $\mathcal{P}_{\rm dw} \gg 1$ the gap is closed. We now explore some astrophysical implications of our results.

\subsection{$M_{\rm gap}$ as a function of $M_*$ and $R$ for passive discs}\label{sec:Mgap_passive}

So far we assumed a fixed stellar mass, $M_* = 1\msun$, a particular model for the disc aspect ratio $H/R$ vs radius $R$, and a given planet location, $R$.  
To understand broader astrophysical implications of our results, we begin by examining the outer regions of discs, which tend to be passively heated by irradiation from the central star \citep[e.g.,][]{CG97}. We follow \cite{Sinclair20-Gaps} who performed radiative transfer modelling of such discs for a range of protostellar masses and found that the aspect ratio can be approximated as a power-law in both $M_*$ and $R$:
\begin{equation}
    \frac{H}{R} = 0.088\; (R/100AU)^{0.35} (M_*/\msun)^{-0.425}.
    \label{HtoR_Sinclair}
\end{equation}

This equation predicts a very weak dependence of the gap opening mass $M_{\rm gap}$ on $M_*$ at a fixed radius $R$. This can be seen as following. The gap opening planet mass to star mass ratio, $q_{\rm gap}$, is obtained by solving the condition $\mathcal{P}_{\rm dw} = 1$ in eq.~(\ref{Crida_mod}). We observe that $q_{\rm dw}$ is proportional to either second or third power of $h = (H/R)$. The former case is realised in weakly turbulent discs, when either the first or the last terms in eq.~(\ref{Crida_mod}) dominate, whereas the latter case is realised for larger $\alpha_{\rm v}$ when the middle term in eq.~(\ref{Crida_mod}) dominates. We therefore obtain
\begin{equation}
    M_{\rm gap} = M_* q_{\rm dw} \propto
    \begin{cases}
    M_*^{0.15}, \quad \text{if }\; q_{\rm dw}\propto h^2\\
    M_*^{-0.275}, \quad \text{if }\; q_{\rm dw}\propto h^3\;.
    \end{cases}
    \label{Mgap_vs_M*}
\end{equation}
These two power-law indexes are both rather shallow.

\begin{figure}
\begin{centering}
\includegraphics[width=1\columnwidth]{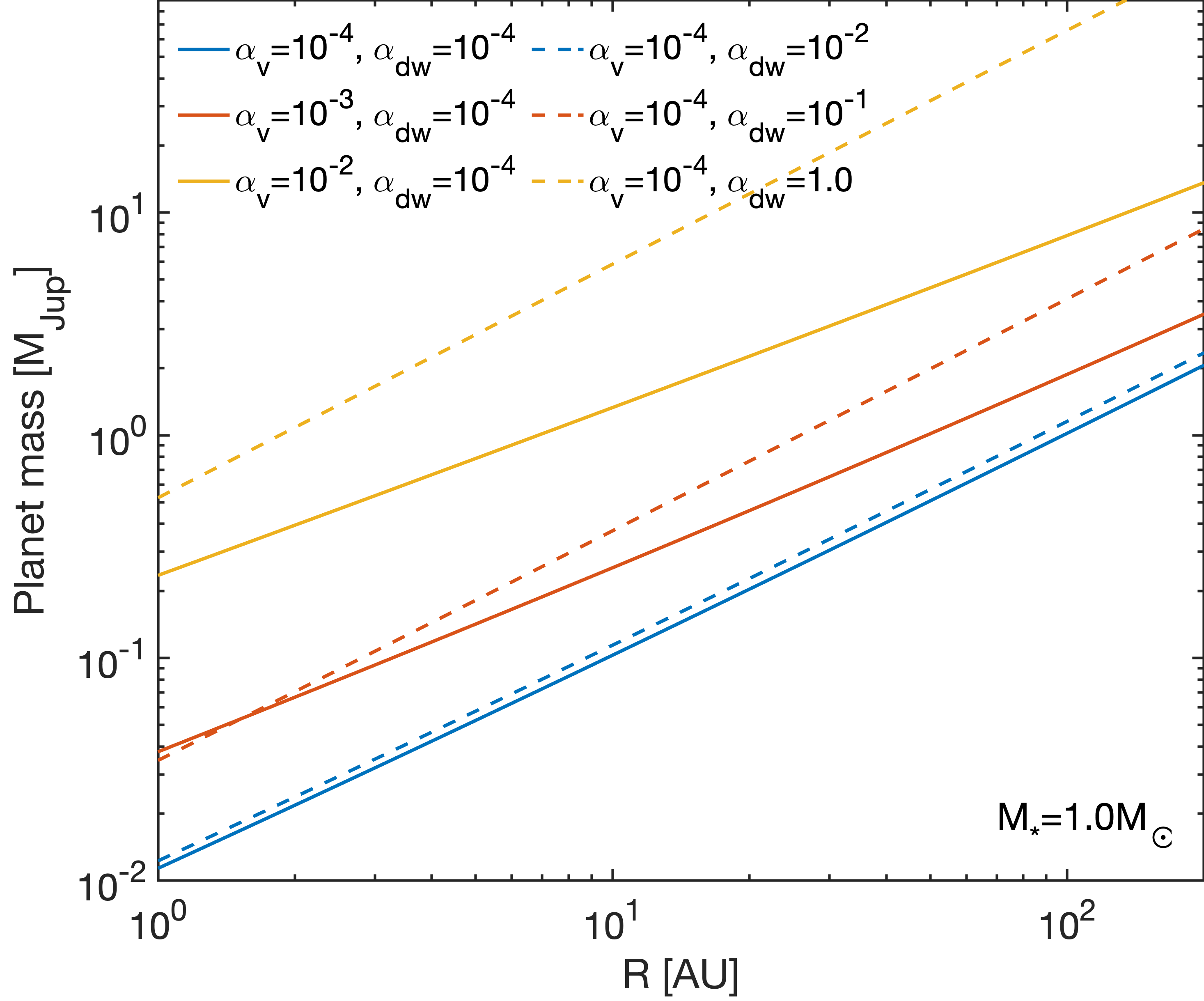}
\par\end{centering}
\caption{\label{fig:Mgap_alpha} Radial dependence of the gas gap opening planet mass $M_{\rm gap}$ for discs around a star with mass $M_*=1\msun$ for a range of turbulent and disc wind parameters. The solid curves are cases with a low fixed $\alpha_{\rm dw}$ values and a varying $\alpha_{\rm v}$, whereas the dashed curves are those with low $\alpha_{\rm v} = 10^{-4}$ but varying $\alpha_{\rm dw}$.}
\end{figure}

Fig.~\ref{fig:Mgap_alpha} shows the radial dependence of $M_{\rm gap}$ obtained using eq.~(\ref{HtoR_Sinclair}) for $H/R$ and solving the equation $\mathcal{P}_{\rm dw} = 1$ for $M_* = 1\msun$. A range of values for $\alpha_{\rm v}$ and $\alpha_{\rm dw}$ is considered in the figure. The solid lines are for fixed $\alpha_{\rm dw}$ but different values of $\alpha_{\rm v}$, whereas the dashed lines all have $\alpha_{\rm v} = 10^{-4}$ but different $\alpha_{\rm dw}$. In all the cases we see that $M_{\rm gap}$ is a strongly increasing function of $R$. The solid blue line can be called the baseline $M_{\rm gap}$; it shows the case of both weak turbulence and weak MHD disc winds, $\alpha_{\rm v} = \alpha_{\rm dw} = 10^{-4}$. In discs with weak to moderate MHD disc winds ($\alpha_{\rm dw} \lesssim 10^{-2}$), turbulent viscosity dominates the gap closing process. In order to significantly exceed the pressure-only based gap opening mass (the solid blue curve), the disc turbulence needs to exceed $\alpha_{\rm v} \sim 10^{-3}$, which is close to the upper limit for many of the discs in which disc turbulence was characterised by ALMA (for references see Introduction). For $M_{\rm gap}$ to be of order a few Jupiter masses at tens of AU we need larger $\alpha_{\rm v}$.

Fig. \ref{fig:Mgap_alpha} shows that, for weak turbulence, $M_{\rm gap}$ would exceed a few Jupiter masses only if $\alpha_{\rm dw} \gtrsim 0.1$. Such high values of $\alpha_{\rm dw}$ are not ruled out unambiguously by ALMA observations of turbulence, however they are unlikely based on the disc evolution perspective. Let us define 
\begin{equation}
    t_{\rm dw} = \frac{R}{|v_{\rm dw}|} \approx 10^6 \left[\frac{R}{100\mathrm{AU}}\right]^{0.8} \left[\frac{\alpha_{\mathrm{dw}}}{0.01}\right]^{-1} \left[\frac{M_*}{\msun}\right]^{-1.35}\; \mathrm{yr}
    \label{tdw}
\end{equation}
Here we used eq.~(\ref{HtoR_Sinclair}) for $H/R$.  

\bref{We note that in our study we neglect the mass loss from the disc due to the MHD and photoevaporative winds in eq. \ref{dSigma_dt}. However, both of these phenomena are essential processes that controls the lifetime of protoplanetary discs \citep[e.g.][]{ClarkeEtal01a, 2020KunitomoInutsuka}. Consideration of the mass loss sets constraints on the upper value of $\alpha_{\rm dw}$ and/or $\alpha_{\rm v}$ for astrophysical discs.}
Given that many of the observed protoplanetary discs survive up to a few Myr, we require $\alpha_{\rm dw} \leq 3\times 10^{-3}$ at least in the time-averaged sense. For higher values of $\alpha_{\rm dw}$ -- and/or higher values of $\alpha_{\rm v}$ -- the gas disc material would be depleted by gas accretion onto the star too rapidly.

We therefore conclude that for realistic passive astrophysical discs, that is, those constrained by ALMA turbulence observations to $\alpha_{\rm v} \lesssim 10^{-3}$ and by the disc lifetime condition (eq.~(\ref{tdw})), the gap opening mass is below $0.1\mj$ inside 1 AU, and exceeds $1 \mj$ only beyond $\sim 100$~AU. 



\begin{figure}
\begin{centering}
\includegraphics[width=1\columnwidth]{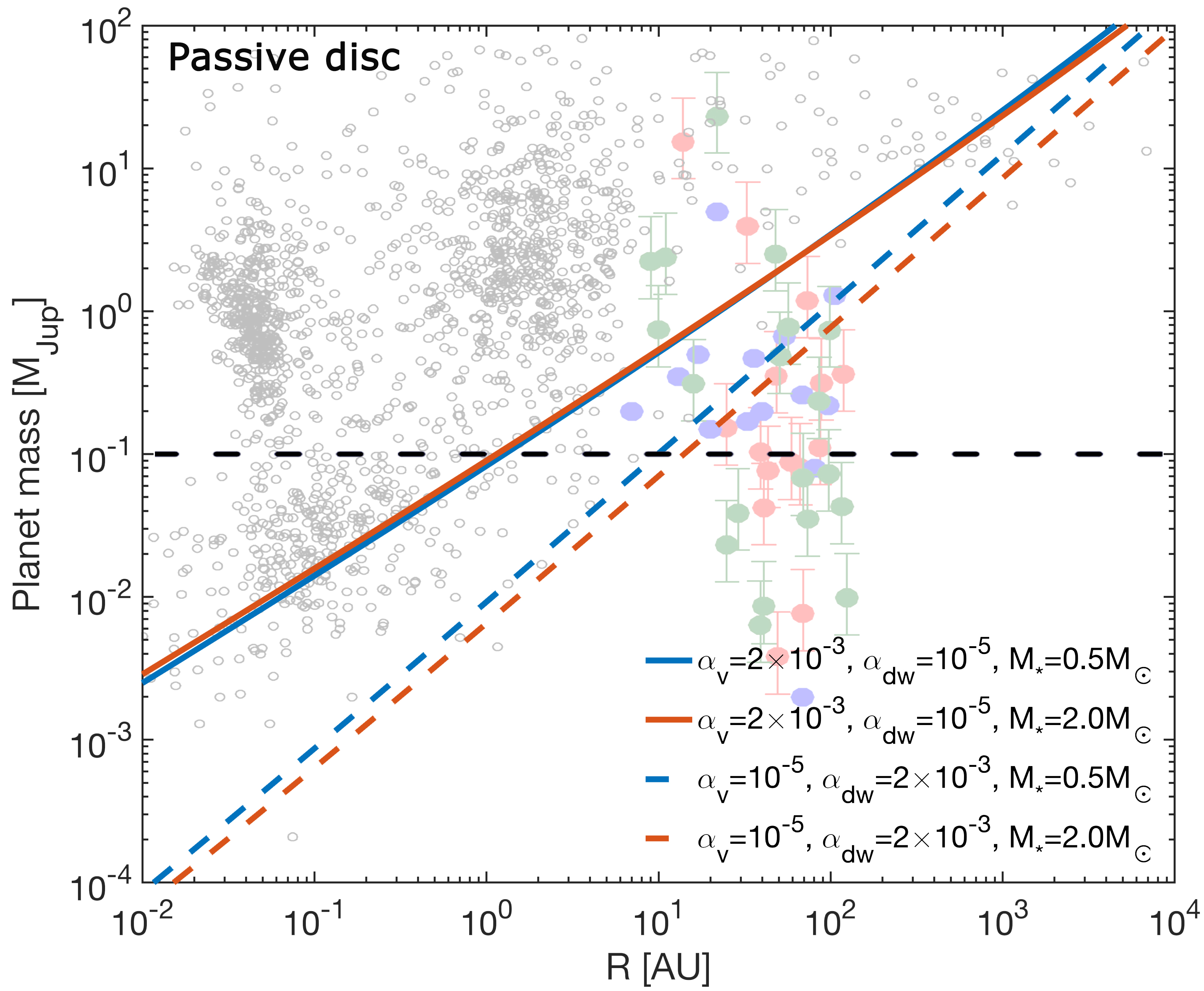}
\par\end{centering}
\caption{\label{fig:Mgap} Similar to fig. \ref{fig:Mgap_alpha} but for stellar masses of 0.5$\msun$ and 2$\msun$ overlaid on the candidate planet mass vs $R$ diagram from \citet{LodatoEtal19} (see text for detail). 
The solid and dashed lines show discs with majority viscous ($\alpha_{\rm v} = 2\times 10^{-3}$) or majority MHD disc wind  ($\alpha_{\rm dw} = 2\times 10^{-3}$) mass transfer, respectively.} 
\end{figure}

In Fig.~\ref{fig:Mgap} we show the radial dependence of $M_{\rm gap}$, similar to the ones shown in Fig.~\ref{fig:Mgap_alpha}, but for the stars with $M_* = 0.5\msun$ (blue lines) and $M_* = 2\msun$ (red lines). In this figure, we picked two opposing limiting cases of angular momentum transfer. 
The solid lines in Fig.~\ref{fig:Mgap} show the case of discs with weak MHD winds and strong turbulence, which we operationally define as $\alpha_{\rm v} = 2\times 10^{-3}$. This value of $\alpha_{\rm v}$ is consistent with the observational upper limits on the disc turbulence, as explained in the Introduction, and is also used in
\cite{Bern20-1} in a recent population synthesis of planet formation. The opposite case of discs with strong MHD disc wind ($\alpha_{\rm dw} = 2\times 10^{-3}$) and weak turbulence is shown with the blue and red dashed lines. The black horizontal dashed line marks the approximate protoplanetary mass at which the runaway growth phase starts in the core accretion scenario.\footnote{Note that earlier calculations suggested that the total gas runaway planet mass (the core plus the gas envelope) is about $0.07 \mj$ \citep[e.g.,][]{Ikoma-00-Mcrit}, whereas more recently authors favour values as large as $\sim 0.2\mj$ \citep[see fig. 16 in][]{2021Emsenhuber}. This appears due to several factors, amongst them ALMA observations requiring earlier massive core formation, thus higher solid accretion rates onto the cores; whereas simulations \citep{OrmelEtal15a,LambrechtsLega17,LambrechtsEtal19-No_CA-runaway}, and microlensing and ALMA observations all indicate lower gas accretion rates onto massive cores \citep{SuzukiEtal18,NayakshinEtal19,Bennet21-No-runaway-desert}.}
The background image in Fig. \ref{fig:Mgap} shows the compilation of candidate planet masses  versus separation from their host stars from \cite{LodatoEtal19}. In this compilation, the empty black circles are the exoplanets detected by radial velocity, transit, microlensing or direct imaging methods, taken from \textit{exoplanets.org}. These planets orbit older stars that lost their protoplanetary discs. The coloured points are the planets around protostars with discs still present; red points taken from \cite{2018LongPinilla}, green from \cite{2018ZhangDSHARP}, and blue are from \cite{BaeEtal18-jupiters}. 

We see from  Fig. \ref{fig:Mgap} that the dependence of the gap opening planet mass on $M_*$ is rather weak in both strong and weak magnetised wind cases, and therefore we can simply neglect it henceforth.

\subsection{$M_{\rm gap}$ in actively accreting discs}\label{sec:Mgap_active}

In the previous section, we studied passive discs, those heated by irradiation from their star. We now consider an example of an actively accreting disc, where internal disc heating may be the dominant heating source.

The midplane disc temperature for active discs is calculated assuming the vertical energy balance of the standard viscously heated and radiatively cooled accretion discs \citep{Shakura73}:
\begin{equation}
\frac{9}{8}\nu\Sigma\Omega^2 = \frac{\sigma_{\mathrm{B}} T^4}{\tau + \tau^{-1}} 
\label{Tvisc}
\end{equation}
where $\tau=\kappa(\rho,T)\rho H$ is the disc optical depth, $\rho$ and $\kappa$ are the midplane gas density and the opacity coefficient, respectively. We use protoplanetary disc opacity from \cite{2012ZhuHartmann}. Further, in the steady state disc, $3\pi \nu \Sigma = \dot M [1 - \sqrt{r_{\rm in}/r}] $, where $r_{\rm in} = 0.01$~AU is the inner boundary of the disc. Eq.~(\ref{Tvisc}) is solved iteratively together with the hydrostatic balance equation; this is necessary since the gas mean molecular weight is a function of $\rho$ and $T$ through the equation of state for the disc.

\begin{figure}
\begin{centering}
\includegraphics[width=1\columnwidth]{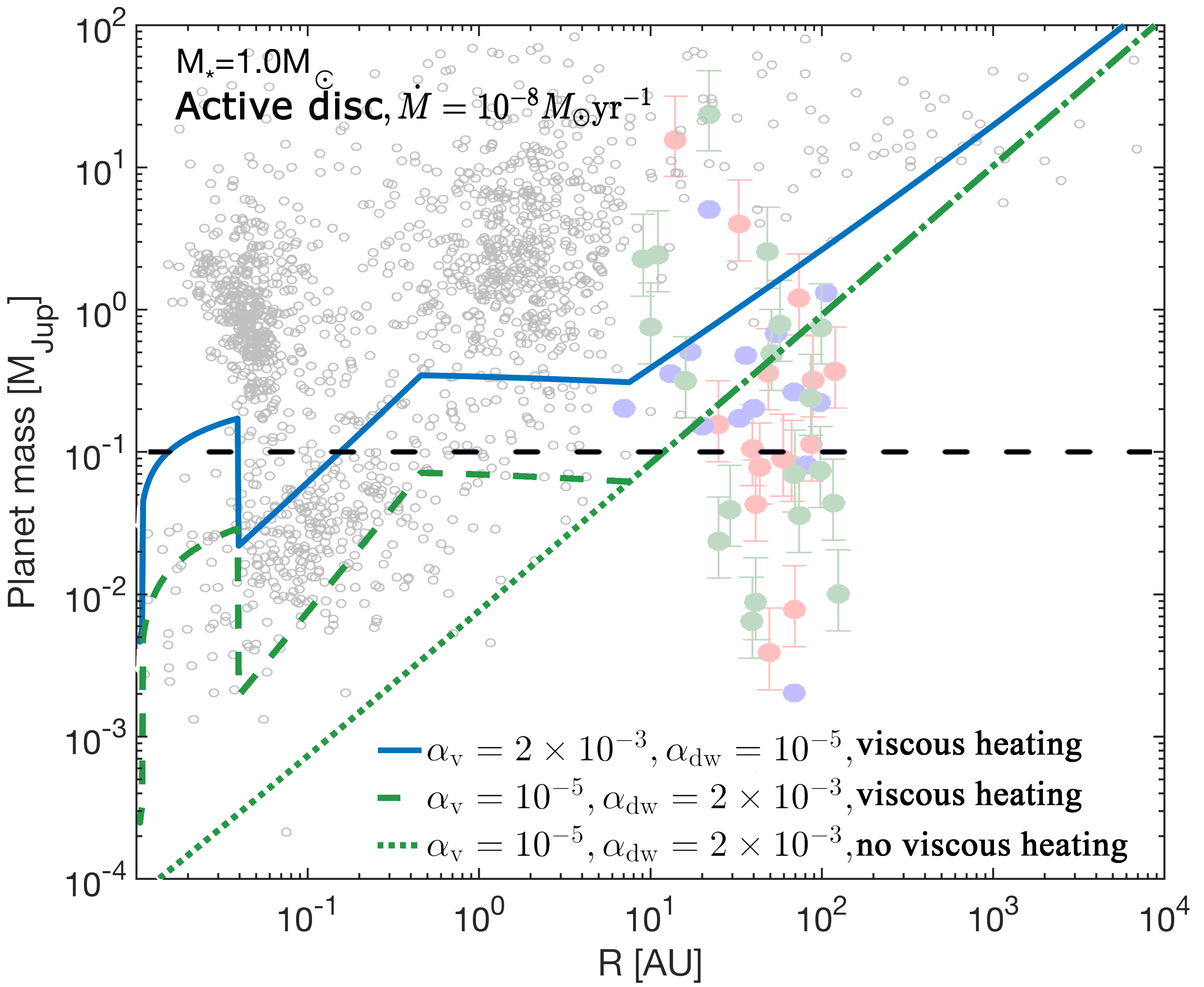}
\par\end{centering}
\caption{\label{fig:Mgap_right} Similar to Fig. \ref{fig:Mgap}, but for a disc around $M_*=1\msun$ star accreting gas at accretion rate $\dot M = 10^{-8} \msun$~year$^{-1}$. The dashed horizontal line shows the approximate planet mass at which the gas accretion runaway is expected to occur in the core accretion scenario.} 
\end{figure}

Fig. \ref{fig:Mgap_right} shows the same information as Fig. \ref{fig:Mgap} but for an actively accreting disc with accretion rate of $\dot M = 10^{-8}\, \msun$~yr$^{-1}$ for $M_*=1\msun$. As in Fig. \ref{fig:Mgap}, we contrast here the case of a disc dominated by turbulent viscosity transport (the solid curve) with the case of a disc powered by a magnetised wind (dashed and dotted curves). All the difference between Figs.  \ref{fig:Mgap_right} and \ref{fig:Mgap} is in the value of $H/R$ as a function of radius used when solving for $M_{\rm gap}$ (which is done by setting $\mathcal{P}_{\rm dw} = 1$ in eq.~(\ref{Crida_mod})).
For the viscosity dominated case, we use $H/R$ for the steady-state Shakura-Sunyaev disc solution at the accretion rate $\dot M$ obtained as described above. In the regions of the disc outward of $R\approx 10$~AU, these $H/R$ values are smaller than that of the passive disc given by eq.~(\ref{HtoR_Sinclair}). This is expected as for any $\dot M$ there is an outer region of the disc heated mainly by irradiation from the star rather than the internal viscous heating. In other words, for $\dot M = 10^{-8} \msun$~year$^{-1}$, the inner 10 AU of the disc are actively heated; the disc outside this region is passively heated.
As a result, for $R< 10$AU, $M_{\rm gap}$ shown in Fig. \ref{fig:Mgap_right} are significantly larger than they are in Fig. \ref{fig:Mgap}. In the regions exterior to 10 AU, $M_{\rm gap}$ is the same in both figures.

Note that  $M_{\rm gap}$ of a  viscously heated disc in Fig. \ref{fig:Mgap_right} shows several transitions in behaviour with $R$, and is in fact non-monotonic. This is due to disc opacity transitions, such as water ice or metallic grains vaporising at certain temperatures. Most prominent of such $M_{\rm gap}$ deviations from monotinic trends with $R$ is the ``inversion" at $R \sim 0.04$~AU, where the disc midplane temperature makes a sharp transition from $\sim 10^3$~K outside of this radius to $\sim 10^4$~K inside. 

For the magnetised wind dominated case, we show two cases in Fig. \ref{fig:Mgap_right} due to physical uncertainties in the energy equation for such discs. As is well known \citep[e.g.,][]{Shakura73}, in purely turbulent viscous discs, the energy is assumed to be generated mainly in the disc midplane and is carried away to infinity by radiation. In the case of MHD disc winds, a significant fraction of the energy can be carried away by the wind itself \citep[e.g., see][and references there]{Suzuki16-MHD-winds}. Further, \cite{2019Mori} demonstrate with non-ideal MHD simulations that most of the heating in such discs occur not in the disc midplane but at a height of several scalehights $H$ above the disc. Since the escape route for radiation in this case is much shorter, these discs are much cooler than MRI-turbulent discs. In effect, the temperature structure of such discs is close to that of passive discs despite significant accretion taking place. This is a topic of ongoing research with answers depending on the geometry of the magnetic fields. 

Here we chose to only bracket the uncertainties in the temperature structure of discs with magnetised winds. The green dashed curve in Fig. \ref{fig:Mgap_right} uses the same $H/R$ as in the purely viscous case (i.e., same as used to compute the blue curve). This curve is appropriate for a hypothetical case where all of the angular momentum removal is due to magnetised winds emanating from the upper layers of the disc, but the heating is still  all released deep in the disc midplane. The green dotted curve is the opposite extreme in which we neglected by the heating generated by MHD disc winds, and we set  $H/R$ to that given by stellar irradiation only, eq.~(\ref{HtoR_Sinclair}). Both of these two extremes are unlikely to be realised in real systems, which we expect to be positioned in-between the extremes, and most probably closer to the dotted one, given the results of \cite{2019Mori}.

\subsection{Implications for the observed planetary populations}\label{sec:Mgap_vs_obs}

Protoplanetary discs are expected to go through different phases in their evolution, beginning with a high mass, high accretion rate phase, proceeding to a lower mass and accretion rate phase before being completely dispersed.
In the previous two sections, we considered the gap opening mass in passive discs and in an active disc with accretion rate $\dot M =10^{-8}\;\msun $~year$^{-1}$. Clearly, these two cases represent just a small part of the parameter phase through which real protoplanetary discs evolve through. Nevertheless, it is possible to make certain observations about the potential impact of magnetised disc winds on planet formation. 


\subsubsection{Formation of gas giants}

As we see from Fig. \ref{fig:Mgap_right}, in the standard turbulent discs $M_{\rm gap}$ may exceed $0.1\mj$ in most of the disc at least while it is actively accreting. This implies that a gas giant planet may be made by gas accretion runaway at any separation from the star. In contrast, for MHD wind discs such runaway may not occur in the disc interior to $R\sim 10$~AU since planets detach from the gas disc by opening a deep gap sooner than they can reach the runaway gas accretion state\footnote{We note that mass flow rate through the gap is significantly lower (by a factor of 4-10) than the mass flow rate outside the gap \citep[e.g.,][]{2006LubowDAngelo}. Thus, the planet accretes mass at much lower rates and after opening a gap. As a result, the timescale at which gas accretion runaway starts may become longer than the disk lifetime.}. This would imply that gas giant planets in discs with magnetised winds are made in the outer reaches of protoplanetary discs and then migrate inwards. In particular, if mass transfer in astrophysical discs is dominated by magnetised winds, then making hot Jupiters in situ \citep[as suggested by, e.g.,][]{2016Batygin-hot-jupiters} would not be possible at all. On the other hand, massive gas giant planets would be somewhat more likely to survive at very wide orbital separations, e.g., beyond 10 AU because they are more likely to migrate inward in the slower type II rather than type I regime. This applies to planet formation via both Core Accretion and disc fragmentation due to Gravitational Instability of massive discs \citep[e.g., see][for the sensitivity of models to planet migration at large distances]{ForganRice13b,HumphriesEtal19}.

Detailed population synthesis calculations are needed to quantify these effects.

\subsubsection{Super Earths and Sub-Saturn planets}

\textit{Kepler} mission revealed that the most common type of planets are super-Earths, i.e., planets with the masses $1 M_\oplus \lesssim M_{\rm p} \lesssim 10-20 M_\oplus$ \citep[e.g.,][]{2013Petigura_Kepler, 2018ZhuPetrovich_Kepler}. Most of observed super-Earth planets have orbital periods shorter than 100 days  \citep[e.g.,][]{2013Batalha_Kepler}. Formation of this abundant population in the standard Core Accretion theory represents a puzzle since gas accretion is expected to be too efficient at close separations \citep[e.g.,][]{2014Lee-not-jupiters}, so that it is not actually clear why these planets are massive cores with small gaseous envelopes  instead of being massive gas giants.

However, if discs are powered by angular momentum transfer via magnetised winds, then, as we saw in \S\S \ref{sec:Mgap_passive} and \ref{sec:Mgap_active}, super-Earth mass planets may open deep gaps in their discs before they reach the runaway gas accretion threshold mass. These planets are likely to remain sub-Saturn mass. The inner $R \lesssim 10$~AU regions of protoplanetary discs with magnetised winds may thus be engines making super Earth and sub-Saturn planets rather than gas giants. 

On the other hand, while the number of smaller mass planets should be higher (by virtue of avoiding their growth into gas giants)
in discs with MHD winds compared to standard viscous discs, their survival past disc dispersal is another matter. In turbulent discs these planets are migrating via the type I migration regime in most of the disc, and there may be several regions of torque reversal where the planets migrate outwards \citep[e.g.,][]{2008BaruteauMasset,Bitsch13-migration-traps}. This creates planet migration traps that are thought to be very important for retention of planets \citep[e.g., fig. 8 in][]{Bern20-1}. However, in discs with magnetised winds such planets are likely to migrate in the type II regime. The migration time of planets in this regime is given by the viscous time \citep[e.g.,][]{BateEtal03,AA09}, which is short in the inner disc. Population synthesis calculations show that type II migrating planets that reach the inner disc are very likely to be lost into the star \citep{AA09,ColemanNelson14}. Although \cite{Kimming20-wind-driven-migration} found that type II migration planets may migrate outward rather than inward in  discs with magnetised winds, we note that this occurs in their simulations only at very high values of their wind efficiency parameter $b$ that corresponds to our $\alpha_{\rm dw} \gtrsim 0.1$. We argued that such high $\alpha_{\rm dw}$ are unlikely (cf. eq.~(\ref{tdw})). At lower $b$ the planets in \cite{Kimming20-wind-driven-migration} migrate inward as in the standard type II migration. If this result translates into the domain of sub-Neptune mass planets, then their survival chances at sub-AU distances would be significantly lower than in the current viscous disc population synthesis models \citep[e.g.,][]{Bern20-2}. 

How these two opposing effects play out in astrophysical discs must be addressed with detailed future modelling.

 \subsubsection{The masses of ALMA candidate planets at $R\gtrsim 10$~AU}

We concluded from Figs.  \ref{fig:Mgap_alpha} and \ref{fig:Mgap} that smaller mass planets are able to open deep gaps in protoplanetary discs with magnetised winds compared with discs powered by turbulent angular momentum transfer. The difference in $M_{\rm gap}$ depends on the particular values of $\alpha_{\rm v}$ and $\alpha_{\rm dw}$, but at wide separations is typically a factor of a few or more. While dedicated future work is needed to study moderate gap and ring features opened by planets in dusty protoplanetary discs with magnetised winds, it is likely that the masses of ALMA candidate planets in systems compiled by \cite{LodatoEtal19} would have to be revised downward if their discs are dominated by magnetised winds. This may be a welcome outcome given that type I planet migration times deduced for these planets appear to be surprisingly short \citep[e.g.,][]{Nayakshin20-Paradox}, requiring a constant supply of new planets at very wide separation. In the type I migration regime, planet migration time is proportional to the planet mass. If  \cite{LodatoEtal19} planet masses are lower by a factor of a few, then their migration times are correspondingly longer, requiring less frequent giant planet formation in the discs. 

Additionally, \cite{NayakshinEtal19} found that most of the ALMA candidate planets in the \cite{LodatoEtal19} compilation lie in the planet mass desert predicted by the Core Accretion theory \citep[e.g.,][]{IdaLin04b}. When planets reach the gas runaway mass \citep[which we note can be as little as $(10-15) \mearth$ at separations tens of AU wide, see][]{PisoYoudin14}, they gain mass voraciously in the runaway gas accretion process \citep[][]{PollackEtal96,DangeloEtal03}. As they grow through the $\sim$ Saturn mass range in $\sim (0.01-0.1)$~Myr \citep[e.g., see fig. 2 in][]{MordasiniEtal12a}, very few Saturn mass planets are expected to exist per every pre-runaway or Jupiter mass planet; this strongly contradicts the mass function of the candidate ALMA planets \citep[fig. 1 in][]{NayakshinEtal19}. 

It was suggested previously that ALMA candidate planets may have masses lower than those found by \cite{LodatoEtal19} \citep[e.g.,][]{Boley17,DongEtal18-alpha0} if the disc viscosity parameter is sufficiently low, e.g.,  $\alpha_{\rm v}\leq 10^{-4}$. If this were true, most of the ALMA candidate planets could be in the pre-runaway regime, perhaps explaining away  the conundrum of too many Saturns in the mass function. However, for the standard viscously accreting discs such low viscosity parameter values would yield correspondingly low gas accretion rates though the discs, contradicting the substantial gas accretion rates observed for many of the stars hosting ALMA candidate planets \citep[e.g.,][argues for $\alpha_{\rm v} \sim 10^{-2}$ for CI Tau]{ClarkeEtal18}.

The results of our paper suggest a potential solution to this difficulty. If mass and angular momentum transport is mainly due to magnetised disc winds rather than turbulence, then it is possible to have a low disc turbulent viscosity and significant gas accretion onto the star at the same time, provided that $\alpha_{\rm dw} \gtrsim 10^{-3}$. We saw from Fig. \ref{fig:Mgap_alpha} that even at $\alpha_{\rm dw} = 10^{-2}$ the gap opening mass $M_{\rm gap}$ is nearly the same as that at $\alpha_{\rm v} = 10^{-4}$. MHD disc winds may hence resolve the problem of the unexpectedly large population of planets in the runaway desert by bringing the inferred planet masses below the gas runaway mass threshold.

At the same time, the suggestion that ALMA candidate planets are actually lower mass than currently believed is not without a drawback of its own. \cite{Pinte20-Dsharp-Vkinks} used molecular line kinematics to search for localised velocity deviations imposed by planets on gas in protoplanetary discs. Such deviations are a complimentary and a completely independent method for detecting planets embedded in protoplanetary discs \citep[][]{Perez15_v_kinks,Perez18-Co-vkink}. Consistently with other authors \citep[e.g.,][]{Casassus19-HD100546}, \cite{Pinte20-Dsharp-Vkinks} find that the masses of the $\sim O(10)$ candidate planets that they find in the DSHARP sample need to be above $1\mj$ to a few $\mj$, noting that this is on average a factor of $4-10$ larger than the masses found through dust continuum emission modelling \citep[as in the ][sample]{LodatoEtal19}. Lowering the former masses by another factor of a few would lead to a yet larger disagreement between the masses of the {\em same} ALMA candidate planets inferred through the dust disc morphology and gas kinematics, respectively.

Concluding, magnetised disc winds may lead to important implications for the inferred parameters and the expected evolution pathways of ALMA candidate planets. Much more work is needed to quantify detailed ramifications of these effects compared with the standard viscous discs.

\subsection{Outlook for the future gap observations}

ALMA observations revealed that the disc substructures, with the majority of them being axisymmetric, are ubiquitous in bright protoplanetary discs \citep{2018Andrews, 2018HuangAndrews}. Combining  observational data with  hydrodynamical simulations and radiation transfer modelling, it has been shown that the observed annular structures are carved by planets with masses ranging between the mass of Neptune up to $\sim10~M_{\rm Jup}$ and orbital radii $\sim$10-100~au \citep{2018ZhangDSHARP, LodatoEtal19}. The comparative lack of observed planets with mass below Neptune and at orbital distance $<10$~au may be due to the limited angular resolution of the ALMA observations.  Observations with higher angular resolution are needed to observe disc substructures due to super-Earth/terrestrial mass planets at orbital distance $<10$~au. Such observations will become possible with future facilities such as the next-generation Very Large Array \citep[ngVLA,][]{2018Ricci, 2020Harter}, the Square Kilometre Array \citep[SKA,][]{2020Ilee} and the extended ALMA array \citep{2020Carpenter, 2022Burrill}.

As shown in Figs. \ref{fig:Mgap_alpha} and \ref{fig:Mgap}, super-Earth and terrestrial planets are able to open a gap in the gas disc at the radial distance of a few AU. Moreover, as it is shown in Fig.~\ref{fig:5}, a much less massive planet is needed to open a gap in the dust disc than in the gas disc. We therefore predict that a higher number of narrow gaps opened by sub-Neptune mass planets will be observed by the next-generation observational facilities.  As such planets are much more ubiquitous than gas giant planets \citep[e.g.,][]{WF14}, we expect narrow gaps  at closer separation be even more widespread than the comparatively wider gaps observed so far with ALMA beyond 10 AU.



\subsection{Caveats}

We emphasise that our results are preliminary as there remain a number of physical and numerical uncertainties in the structure and evolution of discs governed by magnetised disc winds. We point out just two significant uncertainties here. \cite{McNally20-laminar-3D-migration} perform both 2D and 3D simulations of laminar discs with embedded planets, experimenting with various numerical resolutions and different physical assumptions about thermal structure of their discs. They find that in 3D the amplitude and even the sign of the corotation torque may be different compared with 2D disc models for planets in the embedded (type I) migration regime. The thermal structure of discs with magnetised winds in 3D is a topic of active ongoing research \citep[e.g.,][]{2019Mori}, and it is therefore clear that more work is needed to confirm our conclusions for realistic astrophysical discs.

\section{Conclusions}\label{sec:conclusion}

In this paper, we showed that planets embedded in protoplanetary discs have easier time opening deep gaps in gas and dust if mass and angular momentum transport is dominated by magnetised winds rather than by the standard turbulent viscosity. In brief, this is because transport of material across the gap region of the planet by advection is less efficient in resisting planet tidal torques  compared with that by turbulent diffusion of material in the standard discs. Another way of expressing our results is to say that lower mass planets are able to carve deep gaps in the discs dominated by magnetised winds compared to viscous discs. 

The implications of these results are potentially profound for much of the planet formation parameter space (see Sect.~\ref{sec:Mgap_vs_obs}). Future population synthesis calculations of planet formation, whether in the Core Accretion or in the Gravitational Instability frameworks, must include magnetised disc winds.

\section{Aknowledgement}
\bref{We thank the anonymous referee for a comprehensive report, which helped to improve this paper.} V. E. and S. N. acknowledge the funding from the UK Science and Technologies Facilities Council, grant No. ST/S000453/1. This work made use of the DiRAC Data Intensive service at Leicester, operated by the University of Leicester IT Services, which forms part of the STFC DiRAC HPC Facility (www.dirac.ac.uk). GR acknowledges support from an STFC Ernest Rutherford Fellowship (grant number ST/T003855/1).

\section{Data availability}

The data obtained in our simulations can be made available on reasonable request to the corresponding author.


\bibliographystyle{mnras}
\bibliography{mhd-wind}





\bsp	
\label{lastpage}
\end{document}